\documentclass[journal,10.5pt]{IEEEtran}
\IEEEoverridecommandlockouts
\usepackage{cite}
\usepackage[ruled,vlined]{algorithm2e}
\usepackage{amsmath,amssymb,amsfonts}
\usepackage{algorithmic}
\usepackage{graphicx}
\usepackage{textcomp}
\usepackage{xcolor}
\usepackage{enumitem}
\usepackage{cancel}

\usepackage[none]{hyphenat}

\ifCLASSINFOpdf

\else

\fi

\begin{document}

\title{Channel Estimation for Intelligent Reflecting Surface Assisted MIMO Systems: A Tensor Modeling Approach}

        \author{Gilderlan~T.~de~Ara\'{u}jo,
  Andr\'{e}~L.~F.~de~Almeida, R\'{e}my Boyer
\thanks{Gilderlan T. de Ara\'{u}jo  with Federal Institute of Cear\'{a}, Canind\'{e}, CE, e-mail: gilderlan.tavares@ifce.edu.br.}
\thanks{Andr\'{e} Lima F\'{e}rrer de Almeida is with Wireless Telecommunication Research Group (GTEL), Department of Teleinformatics, Federal University of Cear\'{a}, Fortaleza, CE, e-mail: andre@gtel.ufc.br.}
\thanks{R\'{e}my Boyer with University of Lille-1, CRIStAL Laboratory, France e-mail: remy.boyer@univ-lille.fr.}
\thanks{This work was supported by the Ericsson Research, Sweden, and Ericsson Innovation Center, Brazil, under UFC.48
Technical Cooperation Contract Ericsson/UFC. This study was
financed in part by the Coordenação de Aperfeiçoamento de
Pessoal de Nível Superior - Brasil (CAPES)-Finance Code
001, and CAPES/PRINT Proc. 88887.311965/2018-00. Andr\'{e}~L.~F.~de~Almeida like to acknowledge CNPq for its financial support under the grant 306616/2016-5.}
}


\markboth{Journal of \LaTeX\ Class Files,~Vol.XX, No.~X, XXX}%
{Shell \MakeLowercase{\textit{et al.}}: Bare Demo of IEEEtran.cls for IEEE Journals}

\sloppy

\maketitle

\begin{abstract}
Intelligent reflecting surface (IRS) is an emerging technology for future wireless communications including 5G and especially 6G. It consists of a large 2D array of (semi-)passive scattering elements that control the electromagnetic properties of radio-frequency waves so that the reflected signals add coherently at the intended receiver or destructively to reduce co-channel interference. The promised gains of IRS-assisted communications depend on the accuracy of the channel state information. In this paper, we address the receiver design for an IRS-assisted multiple-input multiple-output (MIMO) communication system \emph{via} a tensor modeling approach aiming at the channel estimation problem using supervised (pilot-assisted) methods. Considering a structured time-domain pattern of pilots and IRS phase shifts, we present two channel estimation methods that rely on a parallel factor (PARAFAC) tensor modeling of the received signals. The first one has a closed-form solution based on a Khatri-Rao factorization of the cascaded MIMO channel, by solving rank-1 matrix approximation problems, while the second one is an iterative alternating estimation scheme. {\color{black} The common feature of both methods is the decoupling of the estimates of the involved MIMO channel matrices (base station-IRS and IRS-user terminal), which provides performance enhancements in comparison to competing methods that are based on unstructured LS estimates of the cascaded channel.} Design recommendations for both methods that guide the choice of the system parameters are discussed. Numerical results show the effectiveness of the proposed receivers, highlight the involved trade-offs, and corroborate their superior performance compared to competing LS-based solutions.
\end{abstract}


\IEEEpeerreviewmaketitle

\section{Introduction}
\label{introduction}
In a typical  wireless propagation environment, the transmitted signals suffer attenuation and scattering caused by absorption and reflection, diffraction, and refraction phenomena. In general, multipath propagation is known as one of the main limiting factors to the performance of a wireless communication system \cite{Singal2010}. Indeed, the randomness of the physical radio environment turns the wireless propagation uncontrollable.

Intelligent reflecting surface (IRS) (also referred to as reconfigurable intelligent surface or software-controlled metasurface) \cite{Basar_2019, Survey_NOVO, Liaskos2018ANW, Jung2019, Huang2019,basar2019reconfigurable} is an emergent and promising technology for future (beyond 5G) wireless communications. It consists  of a 2D array with a large number of passive or semi-passive elements that can control the electromagnetic properties of the radio-frequency waves so that the reflected signals add coherently at the intended receiver or destructively to reduce the co-channel interference. Each element can act independently and can be reconfigured in a software-defined manner by means of an external controller. The IRS does not require dedicated radio-frequency chains and is usually wirelessly powered by an external RF-based source, as opposed to amplify-and-forward or decode-and-forward relays, which require dedicated power sources \cite{Huang2019}.
In the literature, IRS is being considered in a number of application scenarios, such as to provide coverage to users located in a dead zone and to suppress co-channel interference when the user is in the edge of the cell \cite{RUI2019,Zang2020}, and to improve the physical layer security \cite{song2020truly,Secure_MIMO}. Besides, the IRS can be employed for simultaneous wireless information and power transfer in an IoT network \cite{RUI2019}.
Regarding wireless communication systems, recently, \cite{In_GAO_2019}  established a connection between IRS technology and a millimeter wave (mmWave) hybrid MIMO systems. In this case, the authors consider a hybrid MIMO-OFDM assisted by IRS working in the mmWave band.

Recent works have discussed the potentials and challenges of IRS-assisted wireless communications (see, e.g., \cite{Basar_2019}, \cite{Survey_NOVO} and references therein). Among the several open issues, we highlight the acquisition of channel state information. One challenge is related to the assumption that the IRS usually consists of passive elements, which means that the estimation of the cascaded channel should be performed at the receiver based on pilots sent by the transmitter {\em via} the IRS. At this point, the pattern of phase shifts used by the IRS during the training phase plays an important role. In addition, the large number of IRS elements imposes an extra challenge to the address the channel estimation problem. {\color{black} Two approaches have been proposed in the literature. The first one assumes a semi-passive structure, where the IRS has a few active elements connected to receive radio-frequency chains. In this case, the availability of some baseband processing at the IRS facilitates the CSI acquisition. An example of this approach is discussed in \cite{Taha2019}, where the involved channels are estimated by means of compressive sensing. The second approach, which is the one adopted in this paper, assumes a fully passive structure, where the IRS operates by reflecting the impinging waves according to some phase-shift pattern. This is a more challenging scenario, where the estimation of the cascaded (transmitter-IRS-receiver) should be done at the receiver based on pilots sent by the transmitter and reflected by the IRS.}

A few works have addressed the channel estimation problem and provided different solutions to the passive IRS case. In \cite{jensen2019optimal}, a minimum variance unbiased estimator is proposed, and an optimal design of the IRS phase shift matrix is found. The authors of \cite{ZHEN2019} propose a two-stage algorithm by exploiting sparse representations of low-rank multipath channels. In \cite{NING2019}, links between massive MIMO and IRS are discussed in the context of Terahertz communications, and a cooperative channel estimation {\em via} beam training is presented. In \cite{Yaoshen2019}, IRS is proposed as a solution to mitigate the blockage problem in mmWave communications and a channel estimation approach is presented. The work \cite{Jie_CHEN_2019} proposes an uplink channel estimation protocol for an IRS aided multi-user MIMO system applying compressing sensing (CS) methods. In \cite{Jawad_2019}, an IRS-aided MIMO system is considered, and channel estimation is carried out in a two-stage approach, and the IRS-assisted link is estimated by means of an approximate message-passing method. Considering an IRS-assisted internet of things scenario, \cite{S_XIA_2019} formulates a joint active detection and channel estimation based on sparse matrix factorization, matrix completion, and multiple measurement vector problems.

The authors of \cite{C_HU_2019} propose a channel estimation framework where the BS-IRS, IRS-UT, and BS-UT channels are estimated in a two-timescale approach, while in \cite{C_You_2019} a practical transmission protocol is proposed to accomplish channel estimation and passive beamforming. In \cite{nadeem2019}, channel estimation is carried out by resorting to an on-off strategy that sequentially activates the IRS elements one-by-one. The work \cite{Alexander_SAM2020} proposes a parallel factor model to solve the channel estimation problem in a multi-user MISO setting. In general, most of the existing works on IRS-assisted communications consider the multiple input single output (MISO) case, where the receiver station is equipped with a single antenna.

{\color{black}In the last decade, tensor modeling has been successfully applied in a variety of signal processing problems \cite{Cichocki_De_Lathauer_SPMG,Sidiropoulos2017,Yassin_2020,Yassin_Remy,Yassin_Andre,Remy_EUSIPCO}, in particular in the context of wireless communications, involving the design of semi-blind receivers for MIMO systems \cite{CONFAC_Andre,PARATUCK_Andre_Elsevier}, channel estimation methods for cooperative communications \cite{Walter2018,B_SOKAL2020}, direction of arrival estimation and beamforming in array processing \cite{Francesca_And_Pierre,Lucas2019,H_Zhang_ICASSP2020}, and, more recently, compressed channel estimation in massive MIMO systems \cite{DanielCA2019,PauloRGB2019}. As discussed in most of these works, tensor-based signal processing benefits from the powerful uniqueness properties of tensor decompositions while exploiting the multi-dimensional nature of the transmitted/received signals and communication channels.} In this work, we establish an existing connection between IRS-assisted MIMO communications and tensor modeling. By assuming a structured time-domain pattern of pilots and IRS phase shifts, we show that the received signal follows a parallel factor (PARAFAC) tensor model. {\color{black} By exploiting the PARAFAC signal structure in two different ways,} we propose two simple and effective algorithms to estimate the cascaded MIMO channel via decoupling the transmitter-IRS and IRS-receiver MIMO channels, respectively. The first algorithm is a closed-form solution based on the Khatri-Rao factorization {\color{black}(KRF)} of the combined BS-IRS and IRS-UT channels, while the second one consists of an iterative bilinear alternating least squares {\color{black}(BALS)} algorithm. While the first algorithm is a closed-form algebraic and less complex solution, the second one can operate under less restrictive conditions on the system parameters.

The common feature of the two algorithms is that the estimation of the cascaded channel is achieved via decoupling the estimation of the two involved channel matrices, which provides a performance enhancement compared to the direct estimation of the cascaded channel via conventional least squares. {\color{black} By focusing on pilot-assisted channel estimation schemes, this work extends the results of our previous conference paper \cite{Gil_SAM2020} by presenting a more comprehensive formulation of tensor-based IRS-assisted channel estimation methods, while bringing a discussion on the uniqueness conditions for the channel estimation problem considering the proposed receivers, from which useful design recommendations on the training parameters are derived. We discuss how to deal with a nonideal setup where the IRS phase shifts are not perfectly known at the receiver, and provide a solution to handle this problem. In addition, we also present generalizations of the proposed approach to multi-user scenarios. Numerical results corroborate the effectiveness of the proposed channel estimation methods and highlight the involved tradeoffs.}

The contributions of this work are summarized as follows.

\begin{itemize}

    \item Resorting to tensor modeling, we connect the channel estimation problem for IRS-assisted MIMO systems to that of fitting PARAFAC model to a third-order tensor;

    \item We derive two simple pilot-assisted channel estimation algorithms (namely, KRF and BALS) that exploit the algebraic structure of the PARAFAC model of the received signals in two different ways;

    \item We provide system design recommendations for the proposed KRF and BALS receivers that ensure the uniqueness of the channel estimation problem;

    \item {\color{black}  We discuss how to handle perturbations/fluctuations on the IRS phase shifts by means of a joint channel and IRS matrix estimation at the receiver;}

    \item  Generalizations of the proposed tensor signal model to multi-user scenarios is provided, which include the multi-UT and the multi-BS cases;

    \item A detailed derivation of the analytical expressions of the CRB is provided.

\end{itemize}

\

\vspace{-2ex}
\noindent \textit{Notation and properties}: Matrices are represented with boldface capital letters ($\mathbf{A}, \mathbf{B}, \dots)$, and vectors are denoted by boldface lowercase letters ($\mathbf{a}, \mathbf{b}, \dots)$. Tensors are symbolized by calligraphic letters $(\mathcal{A}, \mathcal{B}, \dots)$. Transpose and pseudo-inverse of a matrix $\mathbf{A}$ are denoted as $\mathbf{A}^{\textrm{T}}$ and $\mathbf{A}^\dagger$, respectively. The operator $\textrm{diag}(\mathbf{a})$ forms a diagonal matrix out of its vector argument, while $\ast$, $\circ$, $\diamond$, $\odot$ and $\otimes$ denote the conjugate, outer product, Khatri Rao, Hadamard and Kronecker products, respectively. $\mathbf{I}_N$ denotes the $N \times N$ identity matrix. The operator $\textrm{vec}(\cdot)$ vectorizes an $I \times J$ matrix argument, while $\textrm{unvec}_{I \times J}(\cdot)$ does the opposite operation. Moreover, $\textrm{vecd}(\cdot)$ forms a vector out of the diagonal of its matrix argument. The $n$-mode product between a tensor $\mathcal{Y} \in \mathbb{C}^{I \times J \ldots K}$ and a matrix $\mathbf{A} \in \mathbb{C}^{I \times R}$ is denoted as $\mathcal{X}\times_n\mathbf{A}$, for $1 \leq n \leq N$. An identity $N$-way tensor of dimension $R\times R \cdots \times R$ is denoted as $\mathcal{I}_{N,R}$. {\color{black} The operator $D_i(\mathbf{A})$ forms a diagonal matrix from the $i$-th row of its matrix argument $\mathbf{A}$. } Moreover, $\mathbf{A}_{i.}$ denotes the $i$th row of the matrix $\mathbf{A}$. The operator $\left\lceil{x}\right\rceil$ rounds its fractional argument up to the nearest integer. In this paper, we make use of the following identities:
\begin{equation}
(\mathbf{A} \otimes \mathbf{B})(\mathbf{C} \diamond \mathbf{D}) = (\mathbf{A}\mathbf{C}) \diamond (\mathbf{B}\mathbf{D}).
\label{Propertie Kron x Khatri}
\end{equation}
\begin{equation}
(\mathbf{A} \diamond \mathbf{B})^{\textrm{H}}(\mathbf{C} \diamond \mathbf{D}) = (\mathbf{A}^{\textrm{H}}\mathbf{C}) \odot  (\mathbf{B}^{\textrm{H}}\mathbf{D}).
\label{Propertie Hadmard x Khatri}
\end{equation}
\begin{equation}
\textrm{vec}(\mathbf{A\mathbf{B}\mathbf{C}}) = (\mathbf{C}^{\textrm{T}} \otimes \mathbf{A})\textrm{vec}(\mathbf{B}).
\label{Propertie Vec General}
\end{equation}
\begin{equation}
\textrm{diag}(\mathbf{a})\mathbf{b} = \textrm{diag}(\mathbf{b})\mathbf{a}.
\label{Propertie diag(a)b}
\end{equation}
If $\mathbf{B}$ is a diagonal matrix, we have:
\begin{equation}
\textrm{vec}(\mathbf{A\mathbf{B}\mathbf{C}}) = (\mathbf{C}^{\textrm{T}} \diamond \mathbf{A})\textrm{vecd}(\mathbf{B}).
\label{Propertie Vec restrict}
\end{equation}

\section{System Model}
\label{System_Model}
\begin{figure}[!t]
	\centering\includegraphics[scale=0.28]{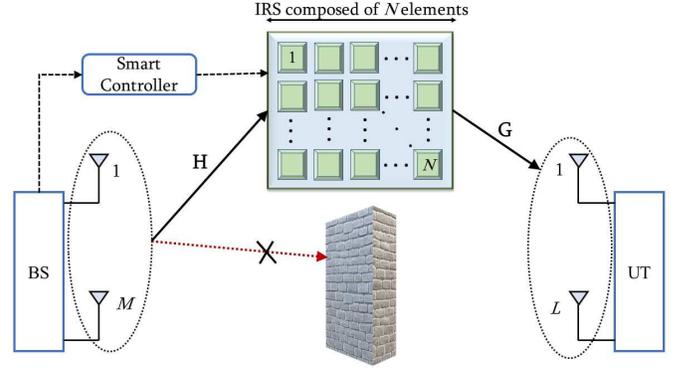}
	\vspace{-2.5ex}
	\caption{IRS-assisted  MIMO system}
	\label{fig:LIS}
\end{figure}

We consider a MIMO communication system assisted by an IRS. Both the transmitter and the receiver are equipped with multiple antennas. Although the terminology adopted in this paper assumes a downlink communication, where the transmitter is the base station (BS) and the receiver is the  user terminal (UT), our signal model also applies to the uplink case by just inverting the roles of the transmitter and the receiver. The BS and UT are equipped with arrays of $M$ and $L$ antennas, respectively. The IRS is composed of $N$ elements, or unit cells, capable of individually adjusting their reflection coefficients (i.e., phase shifts). The system model is illustrated in Figure \ref{fig:LIS}. In a time-slotted transmission, we assume that the IRS adjusts its phase-shifts as a function of the time $t=1, \ldots, T$. We also assume a block-fading channel, which means that the BS-IRS and IRS-UT channels are constant during $T$ time slots. The received signal is given as \cite{ZHEN2019}
\begin{equation}
\mathbf{y}[t] = \mathbf{G}(\mathbf{s}[t] \odot \mathbf{H}\mathbf{x}[t]) + \mathbf{b}[t], \quad 1 \leq  t \leq T,
\label{Eq:EquationFromRef}
\end{equation}
or, alternatively,
\begin{equation}
\mathbf{y}[t] = \mathbf{G}\textrm{diag}(\mathbf{s}[t])\mathbf{H}\mathbf{x}[t] + \mathbf{b}[t],
\label{Eq:EquationFromRef2}
\end{equation}
where $\mathbf{x}[t] \in \mathbb{C}^{M \times 1}$ is the vector containing the transmitted pilot signals at time $t$, $\mathbf{s}[t] = \left[s_{1,t}e^{j\phi_1},\dots,s_{N,t}e^{j\phi_N}\right]^{\textrm{T}} \in \mathbb{C}^{N \times 1}$  is the vector that models the phase shifts and activation pattern of the IRS, where $\phi_n \in (0 , 2\pi]$, and $s_{n,t} \in \{0 , 1\}$ controls the on-off state of the corresponding element at time $t$. The matrices $\mathbf{H} \in \mathbb{C}^{N \times M}$ and $\mathbf{G} \in \mathbb{C}^{L \times N}$ denote the BS-IRS and IRS-UT MIMO channels, respectively, while
$\mathbf{b}[t] \in \mathbb{C}^{L \times 1}$ is the additive white Gaussian noise (AWGN) vector.

The channel {\color{black}training} time $T_s$ is divided into $K$ blocks, where each block has $T$ time slots so that $T_s = KT$. Let us define $\mathbf{y}[k,t]\doteq y[(k-1)T + t]$ as the received signal at the $t$-th time slot of the $k$-th block, $t=1,\dots, T$, $k=1, \dots, K$. Likewise, denote $\mathbf{x}[k,t]$ and $\mathbf{s}[k,t]$ as the pilot signal and phase shift vectors associated with the $t$-th time slot of the $k$-th block. We propose the following structured time-domain protocol: i) the IRS phase shift vector is constant during the $T$ time slots of the $k$-th block and varies from block to block; ii) the pilot signals $\{\mathbf{x}[1], \dots, \mathbf{x}[T]\}$ are repeated over the $K$ blocks. Mathematically, this means that
\begin{eqnarray}
&\mathbf{s}[k,t] = \mathbf{s}[k], \,\, \text{for}\,\, 1\leq t\leq T,\\
&\mathbf{x}[k,t] = \mathbf{x}[t], \,\, \text{for}\,\, 1\leq k \leq K.
\end{eqnarray}

\begin{figure}[!t]
	\centering\includegraphics[scale=0.88]{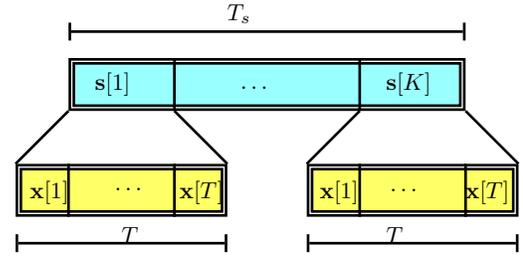}
	\vspace{-2ex}
	\caption{Structured pilot pattern in the time domain}
	\label{fig:bockdiagram}
\end{figure}

An illustration of this time-domain protocol is shown in Figure \ref{fig:bockdiagram}. Under these assumptions, the received signal model (\ref{Eq:EquationFromRef2}) can be written as
\label{sysmodel}
\begin{equation}
\mathbf{y}[k,t] = \mathbf{G}\textrm{diag}(\mathbf{s}[k])\mathbf{H}\mathbf{x}[t] + \mathbf{b}[k,t].
\label{Eq:ReceivedVector}
\end{equation}

Collecting the received signals during $T$ time slots for the $k$-th block in $\mathbf{Y}[k] = \left[\mathbf{y}[k,1] \dots \mathbf{y}[k,T]\right] \in \mathbf{C}^{L \times T}$ leads to
\begin{equation}
\mathbf{Y}[k] = \mathbf{G}\textrm{diag}(\mathbf{s}[k])\mathbf{H}\mathbf{X}^\textrm{T} + \mathbf{B}[k],
\label{Eq:PARAFAC_MODEL}
\end{equation}
where $\mathbf{X}\doteq [\mathbf{x}[1], \dots, \mathbf{x}[T]]^{\textrm{T}} \, \in \mathbb{C}^{T \times M}$, and $\mathbf{B}\doteq [\mathbf{b}[1], \dots, \mathbf{b}[T]]^{\textrm{T}} \, \in \mathbb{C}^{L \times T}$.

\subsection{Least squares channel estimation}
A baseline method consists of estimating a combined version of the communication channels $\mathbf{G}$ and $\mathbf{H}$ using least squares (LS) approach. To derive the LS estimator, we apply property  (\ref{Propertie Vec restrict}) to (\ref{Eq:PARAFAC_MODEL}) to obtain
\begin{eqnarray}
     \mathbf{y}[k] & = & \left(\mathbf{XH^\textrm{T}} \diamond \mathbf{G} \right)\mathbf{s}_k\nonumber\\
     & = & \left(\mathbf{X} \otimes \mathbf{I}_L\right)\left(\mathbf{H}^\textrm{T} \diamond \mathbf{G} \right)\mathbf{s}[k] + \mathbf{b}[k],
    \label{LS_2}
\end{eqnarray}
where $\mathbf{y}[k]\doteq \textrm{vec}(\mathbf{Y}[k]) \in \mathbb{C}^{LT}$, $\mathbf{b}[k]\doteq \textrm{vec}(\mathbf{B}[k]) \in \mathbb{C}^{LT}$, and we have used property (\ref{Propertie Kron x Khatri}). Defining $\tilde{\mathbf{Y}} \doteq \left[\mathbf{y}[1] \dots \mathbf{y}[K]\right] \in \mathbb{C}^{LT \times K}$, and  $\tilde{\mathbf{X}} \doteq \left(\mathbf{X} \otimes \mathbf{I}_L\right) \in \mathbb{C}^{TL \times ML}$, we have
\begin{equation}
     \tilde{\mathbf{Y}}  = \tilde{\mathbf{X}} \left(\mathbf{H}^\textrm{T} \diamond \mathbf{G} \right)\mathbf{S}^{\textrm{T}} + \mathbf{B},
    \label{LS_3}
\end{equation}
where $\mathbf{S}\doteq [\mathbf{s}[1], \ldots, \mathbf{s}[K]]^{\textrm{T}} \in \mathbb{C}^{K \times N}$, $\mathbf{B} \in \mathbb{C}^{LT \times K}$ is the noise matrix constructed in the same way as $\mathbf{Y}$. Finally, defining $\mathbf{y}\doteq \textrm{vec}(\tilde{\mathbf{Y}}) \in \mathbb{C}^{LTK}$, and applying property (\ref{Propertie Vec General}) to (\ref{LS_3}), we get
\begin{equation}
    \mathbf{y} = (\mathbf{S} \otimes  \tilde{\mathbf{X}})\textrm{vec}\left(\mathbf{H^\textrm{T}} \diamond \mathbf{G}\right) + \mathbf{b},
    \label{LS_4}
\end{equation}
or, compactly,
\begin{equation}
     \mathbf{y} = \mathbf{U}\boldsymbol{\theta} + \mathbf{b},
    \label{Y3final}
\end{equation}
where $\mathbf{U} \doteq \mathbf{S} \otimes \tilde{\mathbf{X}} \in \mathbb{C}^{KTL \times NML}$ and $\boldsymbol{\theta} \doteq \textrm{vec}\left(\mathbf{H}^{\textrm{T}} \diamond \mathbf{G}\right)  \in \mathbb{C}^{MLN}$ is the \textit{composite channel parameter} combining the BS-IRS and IRS-UT channels. The LS estimate of this composite channel minimizes the problem
\begin{equation}
\hat{\boldsymbol{\theta}} = \underset{\boldsymbol{\theta}}{\arg\min} \,\, \left\|\mathbf{y} - \mathbf{U}\boldsymbol{\theta}\right\|^2,\label{eq:convLS}
\end{equation}
the solution of which is known to be $\boldsymbol{\theta}=\mathbf{U}^{\dagger}\mathbf{y}$. The computation of this solution can indeed be simplified to $\boldsymbol{\theta}=(\mathbf{S}^{\dagger}\otimes \tilde{\mathbf{X}}^{\dagger})\mathbf{y}$, due to the Kronecker structure of $\mathbf{U}$.

It should be noted that the conventional LS problem ignores the Katri-Rao structure of the composite channel that is present in the linearized parameter vector $\boldsymbol{\theta}$. Indeed, the signal model (\ref{Eq:PARAFAC_MODEL}), or equivalently, (\ref{LS_3}) has a tensor structure, and can be recast as a PARAFAC tensor model. {\color{black} As we show in Section~\ref{Chanel estmation}, adopting this tensor modeling allows us to enhance the channel estimation accuracy  (compared to conventional LS methods). This is achieved by decoupling the estimates of $\mathbf{H}$ and $\mathbf{G}$, rather them estimating $\boldsymbol{\theta}= \textrm{vec}\left(\mathbf{H}^{\textrm{T}} \diamond \mathbf{G}\right)$ as a whole. Moreover, useful system design recommendations can be derived from the proposed modeling approach.}

\subsection{Tensor signal modeling}\label{eq:tensor_modeling}
In order to simplify the exposition of the signal model, we remove the noise term from the following developments. The noise term will be taken into account later. We can rewrite the signal part of equation (\ref{Eq:PARAFAC_MODEL}) as
\begin{equation}
\overline{\mathbf{Y}}[k] = \mathbf{G}\mathbf{D}_k(\mathbf{S}) \mathbf{Z}^\textrm{T}, \quad \mathbf{Z} \doteq \mathbf{X}\mathbf{H}^\textrm{T} \in \mathbb{C}^{T \times N},
\label{eq:PARAFAC MODEL2}	
\end{equation}
where $\mathbf{D}_k(\mathbf{S}) \doteq \textrm{diag}(\mathbf{s}[k])$ denotes a diagonal matrix holding the $k$-th row of the IRS phase shift matrix $\mathbf{S}$ on its main diagonal. The matrix $\overline{\mathbf{Y}}[k]$ can be viewed as the $k$-th frontal matrix slice of a three-way tensor $\overline{\mathcal{Y}} \in \mathbb{C}^{L \times T \times K}$ that follows a PARAFAC decomposition, also known as canonical polyadic decomposition (CPD) \cite{Harshman70,Kolda_TDA,CLdA09,Almeida2016,Sidiropoulos2017}). Each $(\ell,t,k)$-th entry of the noiseless received signal tensor $\overline{\mathcal{Y}}$ can be written as:
\begin{equation}\label{scalar_Y}
\overline{y}_{\ell,t,k}= \sum\limits_{n=1}^N g_{\ell,n}z_{t,n}s_{k,n},
\end{equation}
where $g_{\ell,n}\doteq [\mathbf{G}]_{\ell,n}$, $z_{t,n}\doteq [\mathbf{Z}]_{t,n}$, and $s_{k,n}\doteq [\mathbf{S}]_{k,n}$. A shorthand notation for the PARAFAC decomposition (\ref{scalar_Y}) is denoted as $\overline{\mathcal{Y}}=[[\mathbf{G}, \mathbf{Z}, \mathbf{S}]]$. Using the $n$-mode product notation, the PARAFAC decomposition of the noiseless received signal tensor $\overline{\mathcal{Y}}$ can be represented as
\begin{eqnarray}
    \overline{\mathcal{Y}} & = &\mathcal{I}_{3,N} \times_1 \mathbf{G} \times_2 \mathbf{Z} \times_3 \mathbf{S}.
    \label{n-mode notation}
\end{eqnarray}
Exploiting the trilinearity of the PARAFAC decomposition, we can ``unfold'' received signal tensor $\overline{\mathcal{Y}}$ into the following three matrix forms \cite{Harshman70,Kolda_TDA}:
\begin{eqnarray}
& \overline{\mathbf{Y}}_1= \mathbf{G}(\mathbf{S} \diamond \mathbf{Z})^{\textrm{T}} \,\, \in \mathbb{C}^{L \times TK},\label{unfolding1}\\
&\overline{\mathbf{Y}}_2= \mathbf{Z}(\mathbf{S} \diamond \mathbf{G})^{\textrm{T}} \,\, \in \mathbb{C}^{T \times LK},\label{unfolding2}\\
&\overline{\mathbf{Y}}_3= \mathbf{S}(\mathbf{Z} \diamond \mathbf{G})^{\textrm{T}}\,\, \in \mathbb{C}^{K \times LT},\label{unfolding3}
\end{eqnarray}
where $\overline{\mathbf{Y}}_1\doteq [\overline{\mathbf{Y}}[1], \dots, \overline{\mathbf{Y}}[K]]$, $\overline{\mathbf{Y}}_2\doteq [\overline{\mathbf{Y}}^{\textrm{T}}[1], \dots, \overline{\mathbf{Y}}^{\textrm{T}}[K]]$, and $\overline{\mathbf{Y}}_3\doteq [\textrm{vec}(\overline{\mathbf{Y}}[1]), \dots, \textrm{vec}(\overline{\mathbf{Y}}[K])]^{\textrm{T}}$. In the following, the algebraic structure of the PARAFAC model (\ref{scalar_Y}) is exploited to formulate two channel estimation methods. The PARAFAC model is powerful due to its essential \textcolor{black}{factor identification} uniqueness property, which has its roots on the concept of Kruskal rank (k-rank). Details can be found in \cite{KRUSKAL197795, STEGEMAN2007540}.

\section{Channel Estimation Methods}
\label{Chanel estmation}

Our goal is to estimate the channel matrices $\mathbf{H}$ (BS-IRS) and $\mathbf{G}$ (IRS-UT) from the received signal tensor given in (\ref{scalar_Y}). Let us define $\mathcal{Y} \doteq \overline{\mathcal{Y}} + \mathcal{B}$ as the noise-corrupted received signal tensor, where $\mathcal{B} \in \mathbb{C}^{L \times T \times K}$ is the additive noise tensor. Likewise, $\mathbf{Y}_i\doteq \overline{\mathbf{Y}}_i + \mathbf{B}_i$, $i=1,2,3$, are the noisy versions of the 1-mode, 2-mode and 3-mode matrix unfoldings (\ref{unfolding1})-(\ref{unfolding3}) of the received signal tensor, and $\mathbf{B}_{i=1,2,3}$ the corresponding matrix unfoldings of the noise tensor.

The pilot signal matrix $\mathbf{X}$ and the IRS phase shifts matrix $\mathbf{S}$ can be designed as semi-unitary matrices satisfying $\mathbf{X}^{\textrm{H}}\mathbf{X}=T\mathbf{I}_{M}$ and $\mathbf{S}^{\textrm{H}}\mathbf{S}=K\mathbf{I}_{N}$, respectively. A good choice is to design both $\mathbf{X}$ and $\mathbf{S}$ as truncated discrete Fourier transform (DFT) matrices. The optimal design of the IRS matrix $\mathbf{S}$ is discussed in \cite{jensen2019optimal} for the multiple-input single-output (MISO) case (i.e, for single-antenna users).

\subsection{Khatri-Rao Factorization based channel estimation}
\label{sec:sec5A}
First, note that we can rewrite the noise-corrupted matrix unfolding (\ref{unfolding3}) as:
\begin{equation}
\begin{aligned}
\mathbf{Y}_3  & = \mathbf{S}(\mathbf{Z} \diamond \mathbf{G})^{\textrm{T}} +  \mathbf{B}_3\\
 & = \mathbf{S}\left(\mathbf{H}^{\textrm{T}} \diamond \mathbf{G}\right)^{\textrm{T}}(\mathbf{X}\otimes \mathbf{I}_L)^{\textrm{T}} + \mathbf{B}_3,
\end{aligned}
\label{unfolding3_2}
\end{equation}
where we have applied the property $(\mathbf{A} \otimes \mathbf{B})(\mathbf{C} \diamond \mathbf{D})=(\mathbf{A}\mathbf{C}) \diamond (\mathbf{B}\mathbf{D})$ to the term $(\mathbf{Z} \diamond \mathbf{G})=(\mathbf{X}\mathbf{H}^{\textrm{T}} \diamond \mathbf{G})$.
A bilinear time-domain filtering is applied at the receiver by exploiting the knowledge of the IRS matrix and the pilot signal matrix, as follows
\begin{equation}
\mathbf{\Omega} \doteq (\mathbf{X}^{\dagger}\otimes \mathbf{I}_L)\mathbf{Y}^{\textrm{T}}_3(\mathbf{S}^{\textrm{T}})^{\dagger} = \mathbf{H}^{\textrm{T}} \diamond \mathbf{G} + \widetilde{\mathbf{B}}_3,
\label{Khatri_ZG}
\end{equation}
where $\widetilde{\mathbf{B}}_3=(\mathbf{X}^{\dagger}\otimes \mathbf{I}_L)\mathbf{B}^{\textrm{T}}_3(\mathbf{S}^{\textrm{T}})^{\dagger}$ is the filtered noise term. Note that $\mathbf{\Omega} \in \mathbb{C}^{ML \times N}$ is a noisy version of the (Khatri-Rao structured) \emph{virtual MIMO channel} that models the IRS-assisted MIMO transmission. Due to the semi-unitary structure of $\mathbf{S}$ and $\mathbf{X}$, the correlation properties of the additive noise are not affected by the bilinear filtering step.

From (\ref{Khatri_ZG}), we propose to estimate $\mathbf{H}$ and $\mathbf{G}$ by solving the following Khatri-Rao least squares approximation problem
\begin{equation}
\underset{\mathbf{H}, \mathbf{G}}{\min} \left\| \mathbf{\Omega} - \mathbf{H}^{\textrm{T}} \diamond \mathbf{G}\right\|^2_F.\label{KRF_problem}
\end{equation}
An efficient solution to this problem is given by the Khatri-Rao factorization (KRF) algorithm \cite{Kibangou2009},\cite{Roemer2010}. Note that the problem (\ref{KRF_problem}) can be interpreted as finding estimates of $\mathbf{H}$ and $\mathbf{G}$ that minimize a set of rank-$1$ matrix approximations, i.e.,
\begin{equation}
(\hat{\mathbf{H}},\hat{\mathbf{G}})= \underset{\{\mathbf{h}_n\},\{\mathbf{g}_n\}}{\arg\min}  \sum\limits_{n=1}^N  \left\| \widetilde{\mathbf{\Omega}}_n - \mathbf{g}_n\mathbf{h}^{\textrm{T}}_n \right\|^2_F\label{KRF_problem2},
\end{equation}
where $\widetilde{\mathbf{\Omega}}_n\doteq \textrm{unvec}_{L \times M}(\boldsymbol{\omega}_n) \in \mathbf{C}^{L \times M}$, while $\mathbf{g}_n \in \mathbf{C}^{L \times 1}$ and $\mathbf{h}^{\textrm{T}}_n \in \mathbf{C}^{1 \times M}$ are the $n$-th column of $\mathbf{G}$ and $n$-th row of $\mathbf{H}$, respectively. The estimates of $\mathbf{g}_n$ and $\mathbf{h}_n$ in (\ref{KRF_problem2}) can be obtained from the dominant left and right singular vectors of $\widetilde{\mathbf{\Omega}}_n$, respectively, for $1 \leq n \leq N$. Hence, our channel estimation problem translates into solving $N$ rank-1 matrix approximation subproblems, for which several efficient solutions exist in the literature \cite{golub13}. A summary of the KRF algorithm is given in Algorithm $1$, where t-SVD denotes a truncated SVD (t-SVD denotes also tensor SVD in the tensor literature) that returns the dominant singular vector and its associated singular value.
Once $\hat{\mathbf{H}}$ and $\hat{\mathbf{G}}$ are found from problem (\ref{KRF_problem2}), we can build the composite channel.
\begin{algorithm}[!t]
\IncMargin{1em}
	\DontPrintSemicolon
	\DontPrintSemicolon
	\SetKwData{Left}{left}\SetKwData{This}{this}\SetKwData{Up}{up}
	\SetKwFunction{Union}{Union}\SetKwFunction{FindCompress}{FindCompress}
	\SetKwInOut{Input}{input}\SetKwInOut{Output}{output}
	\textbf{Procedure}\\
	\Output{$\hat{\mathbf{H}}$ and $\hat{\mathbf{G}}$}
	\BlankLine
	\Begin{
			   \textit{Bilinear filtering of $\mathbf{Y}_3$}:\;
			   $\mathbf{\Omega}^{\textrm{T}} \longleftarrow \mathbf{S}^{\textrm{H}}\mathbf{Y}_3(\mathbf{X}^{\ast}\otimes \mathbf{I}_L)$\;
		\For{$n = 1, \dots ,N$}{
			$\widetilde{\mathbf{\Omega}}_n \longleftarrow \textrm{unvec}_{L \times M}(\boldsymbol{\omega}_n)$\;
			$(\mathbf{u}_1,\mathbf{\sigma}_1,\mathbf{v}_1)\longleftarrow\textrm{t-SVD}(\widetilde{\mathbf{\Omega}}_n)$\;
			$\hat{\mathbf{h}}_n \longleftarrow \sqrt{\sigma_1}\mathbf{v}_1^\ast$\;
			$\hat{\mathbf{g}}_n \longleftarrow \sqrt{\sigma_1}\mathbf{u}_1$\;	
			\textbf{end}}
		\textit{Reconstruct} $\hat{\mathbf{H}}$ \textit{and} $\hat{\mathbf{G}}$:\;
		$\hat{\mathbf{H}} \longleftarrow \left[\hat{\mathbf{h}}_1, \dots , \hat{\mathbf{h}}_N\right]^{\textrm{T}}$\;
		$\hat{\mathbf{G}} \longleftarrow \left[\hat{\mathbf{g}}_1, \dots , \hat{\mathbf{g}}_N\right]$\;
		\textbf{end}
	}
	\caption{Khatri-Rao factorization (KRF)
	}
	\label{PseudocodeKRF}
\end{algorithm}

\subsection{BALS channel estimation}\label{sec:bals_parafac}
From the noisy versions of the matrix unfoldings (\ref{unfolding1}) and (\ref{unfolding2}), we can derive an iterative solution based on a bilinear alternating least squares (BALS) algorithm. {\color{black} This algorithm is a simplified version of the well-known trilinear ALS algorithm for estimating the factor matrices of a PARAFAC model \cite{BRO}. In our case, since $\mathbf{S}$ is known at the receiver,} it consists of estimating the matrices $\mathbf{G}$ and $\mathbf{H}$ in an alternating way by iteratively optimizing the following two cost functions:
\begin{eqnarray}
&\hat{\mathbf{G}} = \underset{\mathbf{G}}{\arg\min} \,\, \left\|\mathbf{Y}_{1} - \mathbf{G}(\mathbf{S} \diamond \mathbf{X}\mathbf{H}^{\textrm{T}})^{\textrm{T}}\right\|_F^2\label{func costG},\\
&\hat{\mathbf{H}} = \underset{\mathbf{H}}{\arg\min}\,\, \left\|\mathbf{Y}_{2} - \mathbf{X}\mathbf{H}^{\textrm{T}}(\mathbf{S} \diamond \mathbf{G})^{\textrm{T}}\right\|_F^2,
\label{Func costZ}
\end{eqnarray}
the solutions of which are respectively given by
\begin{eqnarray}
&\hat{\mathbf{G}} = \mathbf{Y}_1\left[\left(\mathbf{S} \diamond \mathbf{X}\mathbf{H}^{\textrm{T}}\right)^{\textrm{T}} \right]^\dagger,\label{EstimaG}\\
&\hat{\mathbf{H}}^{\textrm{T}} = \mathbf{X}^\dagger\mathbf{Y}_2\left[\left(\mathbf{S} \diamond \mathbf{G} \right)^{\textrm{T}}\right]^\dagger. \label{EstimaH}
\end{eqnarray}
\begin{algorithm}[!t]
\IncMargin{1em}
	\DontPrintSemicolon
	\SetKwData{Left}{left}\SetKwData{This}{this}\SetKwData{Up}{up}
	\SetKwFunction{Union}{Union}\SetKwFunction{FindCompress}{FindCompress}
	\SetKwInOut{Input}{input}\SetKwInOut{Output}{output}
	\textbf{Procedure}\\
	\Input{$i = 0$; \textit{Initialize} $\hat{\mathbf{H}}_{(i=0)}$}
	\Output{$\hat{\mathbf{H}}$, $\hat{\mathbf{G}}$} 
	\BlankLine
    \Begin{
	$i = i + 1 ;$\;
     \While{$\|e(i) - e(i-1)\| \geq \delta$}{
	    \begin{enumerate}
		\item [1:]  \textit{Find a least squares estimate of} $\mathbf{G}$:\\
		\vspace{0.1in}
		\begin{itemize}
		    \item[] $\hat{\mathbf{G}}_{(i)} = \mathbf{Y}_1\left[\left(\mathbf{S} \diamond \mathbf{X}\hat{\mathbf{H}}^{\textrm{T}}_{(i-1)}\right)^{\textrm{T}} \right]^\dagger$
		\end{itemize}
		 \vspace{0.1in}
		\item[2:] \textit{Find a least squares estimate of} $\mathbf{H}$:\\
		\vspace{0.1in}
		\begin{itemize}
		    \item[] $\hat{\mathbf{H}}^{\textrm{T}}_{(i)} = \mathbf{X}^\dagger\mathbf{Y}_2\left[\left(\mathbf{S} \diamond \mathbf{G}_{(i)}\right)^{\textrm{T}}\right]^\dagger$
		\end{itemize}
		\vspace{0.1in}
		\item[3:]\textit{Repeat steps} $1$ to $2$ \textit{until convergence.}
	  \end{enumerate}
	  \textbf{end}
	  }
	  \textbf{end}
	  }
	\caption{Bilinear alternating least squares (BALS)}
	\label{PseudocodeBALS}
\end{algorithm}
The convergence is declared when $\|e_{(i)} - e_{(i-1)}\|\leq \delta$, where $e_{(i)} = \|\mathcal{Y} - \hat{\mathcal{Y}}_{(i)}\|_{F}^{2}$ denotes the reconstruction error computed at the $i$-th iteration, $\delta$ is a threshold parameter, and $\hat{\mathcal{Y}}_{(i)}=[\hat{\mathbf{G}}_{(i)}, \mathbf{X}\hat{\mathbf{H}}^{\textrm{T}}_{(i)}, \hat{\mathbf{S}}]$ is the reconstructed PARAFAC model (c.f. (\ref{Eq:PARAFAC_MODEL}), (\ref{scalar_Y})) obtained from the estimated channel matrices $\hat{\mathbf{G}}_{(i)}$ and $\hat{\mathbf{H}}_{(i)}$ at the end of the $i$-th iteration. In this work, we adopt $\epsilon=10^{-5}$. Despite the iterative nature of the BALS algorithm, only a few iterations are necessary for convergence (usually less than 10 iterations) thanks to the knowledge of the IRS matrix $\mathbf{S}$ that remains fixed during the iterations.

If $\mathbf{X}$ and $\mathbf{S}$ are column-orthogonal (which requires $K\geq N$ and $T\geq M$), the right pseudo-inverses in (\ref{EstimaG}) and (\ref{EstimaH}) can be replaced by  matrix products. This leads to a lower complexity implementation of the BALS algorithm with simplified estimation steps, as shown in Appendix B.

\subsection{Computational complexity}
The computational complexity is in general dominated by the (truncated) SVD steps involved in Algorithm 1 (KRF) to compute rank-1 matrix approximations, as well as in Algorithm 2 (BALS) to calculate the LS estimates of the channel matrices. First, recall that computing the SVD of a matrix $P \times Q$ has a complexity order of $\mathcal{O}(PQ\min(P,Q))$, while computing the inner product of two matrices of dimensions $P \times F$ and $F \times Q$ has complexity $\mathcal{O}(PQF)$. The complexity of KRF is that of computing $N$ rank-1 approximation routines from the matrix $\widetilde{\boldsymbol{\Omega}}_n$, $n=1, \ldots, N$, which can be efficiently implemented by means of the well-known power method \cite{golub13}. From these results, we find that the KRF algorithm has a complexity of order $\mathcal{O}(MLN)$ owing to the $N$ rank-$1$ matrix approximations.  As for the iterative BALS receiver, as discussed in Section \ref{sec:bals_parafac}, the computational cost associated with steps $1$ and $2$ is that of computing two right pseudo-inverses per iteration, which is equivalent to $\mathcal{O}(NKT[2N + L])$ and $\mathcal{O}(MT[2M + LK] + N^2[LK + M] + MNLK)$, respectively. Note, however, that this computational cost is greatly reduced when $\mathbf{S}$ and $\mathbf{X}$ have orthogonal columns (which require $K\geq N$ and $T\geq M$, respectively). According to the steps derived in Appendix B, the cost per iteration is that of computing the matrix products in steps $1$ and $2$ of Algorithm \ref{PseudocodeBALS_With K>=N restriction}, which corresponds to $\mathcal{O}(LN[TK + N])$ and $\mathcal{O}(M[LKT + LKN + N^2])$, respectively.

\subsection{Dealing with IRS perturbations and blockages}
In outdoor scenarios, due to the exposure of the IRS to weather and atmospheric conditions, its elements may be subject to unknown blockages, as well as time-dependent fluctuations on their phase and amplitude responses \cite{B_Li_S_Unkown}. Such unknown perturbations have a random nature, and may be caused by water droplets, snowflakes, dry and damp sand particles, among others. In this case, the IRS matrix $\mathbf{S}$ deviates from its desired structure, and the assumption of a perfect knowledge of all the phase shifts at the receiver may not hold. Otherwise stated, the receiver cannot benefit from the full knowledge of the IRS phase shifts to estimate the cascade channel, i.e., it should be able to operate in a \textit{semi-blind} way. Adopting our tensor modeling approach, it is possible to deal with this issue by resorting to a trilinear alternating least squares (TALS) algorithm that jointly estimates $\mathbf{G}$, $\mathbf{H}$, and $\mathbf{S}$ by fully exploiting the trilinear structure of the received signal tensor in (\ref{scalar_Y})-(\ref{n-mode notation}). The TALS algorithm is an extension of the BALS one by adding in Algorithm 2 a third estimation step
$$\hat{\mathbf{S}}_{(i)}= \mathbf{Y}_3\Big[\big( \mathbf{X}\hat{\mathbf{H}}^{\textrm{T}}_{(i)}\diamond \hat{\mathbf{G}}_{(i)}\big)^{\textrm{T}}\Big]^\dagger$$
that includes the update/refinement the IRS matrix within the loop.
The TALS arises as a good alternative to deal with IRS phase shift perturbations. Since the channel matrices are now estimated in a blind way, i.e., without the knowledge of the IRS matrix $\mathbf{S}$, more iterations are required to achieve convergence. Moreover, the complexity is also increased due to the additional LS estimation step at each iteration. TALS is a well-known algorithm for fitting a PARAFAC model \cite{BRO,Kolda_TDA,CLdA09}. In such a blind approach, we can resort to the Kruskal's uniqueness conditions for the PARAFAC model \cite{KRUSKAL197795} to obtain useful system design recommendations. A simplified condition can be obtained when the channel matrices have full rank.\footnote{The condition $\textrm{min}(L,N) + \textrm{min}(M,N) + \textrm{min}(K,N) \geq 2N+2$ usually implies more restrictive choices on the system parameters $L$, $M$, and $K$, compared to the conditions discussed in Section V, which are valid when considering the perfect knowledge of the IRS matrix.} In this case, $\textrm{min}(L,N) + \textrm{min}(M,N) + \textrm{min}(K,N) \geq 2N+2$ guarantees the uniqueness of $\mathbf{G}$, $\mathbf{H}$ and $\mathbf{S}$ (see \cite{STEGEMAN2007540,Sidiropoulos2017} for a deeper uniqueness  discussion in the general case).  {\color{black} It is known that TALS may suffer from slow convergence due to its sensitivity to the initialization. However, several enhancements may be used to improve its performance (see \cite{CLdA09} and references therein).}

\section{Design recommendations}
\label{sec:uniqueness}
The KRF method (Algorithm 1) has a bilinear filtering step as shown in (\ref{Khatri_ZG}) requiring that the IRS phase shift matrix $\mathbf{S}$ and the pilot symbol matrix $\mathbf{X}$ have full column-rank, which implies the following conditions
\begin{equation}
K \geq N \quad \textrm{and} \quad T \geq M. \label{eq:cond_krf}
\end{equation}
As mentioned earlier, a good choice is to design $\mathbf{X}$ and $\mathbf{S}$ are semi-unitary (or column-orthogonal) matrices, for two reasons. First, because the the semi-unitary design replaces matrix inversions in (\ref{Khatri_ZG}) by simple matrix products, simplifying the receiver processing. Second, because the correlation properties of the filtered noise term in (\ref{Khatri_ZG})
are preserved.

The BALS method (Algorithm 2) requires that the two Khatri-Rao product terms $\mathbf{M}_1 = \mathbf{S} \diamond \mathbf{X}\mathbf{H}^{\textrm{T}} \in \mathbb{C}^{KT \times N}$ and $\mathbf{M}_2= \mathbf{S} \diamond \mathbf{G} \in \mathbb{C}^{KL \times N}$ have full column-rank, so that  (\ref{EstimaG}) and (\ref{EstimaH}) (resp. steps 1 and 2 of Algorithm 2) admit unique solutions. This means that the conditions $KT \geq N$ and $KL \geq N$ must be satisfied. Combining these two inequalities implies $\textrm{min}(KT,KL)\geq N$, or, equivalently, $K\textrm{min}(T,L)\geq N$. {\color{black}In addition, the condition $T \geq M$ in (\ref{Func costZ}) is required, since $\mathbf{X}$ must have full column-rank to be left-invertible. Hence, the following conditions are necessary
\begin{equation}
  K\textrm{min}(T,L)\geq N.\quad \textrm{and} \quad T \geq M
  \label{eq:cond_bals}
\end{equation}}

Comparing the conditions (\ref{eq:cond_krf}) and (\ref{eq:cond_bals}), we can note that BALS has a less restrictive requirement on the minimum number $K$ of time blocks for the channel training compared to KRF method. {\color{black} Note that, in the special case $L = 1$ (MISO or SISO system), the inequalities (\ref{eq:cond_krf}) and (\ref{eq:cond_bals}) are identical, i.e., BALS and KRF are subject to the same training requirements. {\color{black} The advantage of BALS over KRF appears in the MIMO case, since BALS can operate under $K<N$, while KRF requires $K\geq N$}\footnote{{\color{black}Note that if $K= 1$, KRF reduces to the conventional LS estimator. However, in this case we cannot resolve/decouple the estimation of the two channel matrices, and the performance gain obtained with such a decoupling \textit{via} solving problem (\ref{KRF_problem}) is lost.}}.} On the other hand, KRF usually has a lower computational complexity than BALS, as will be shown later in the discussion of our numerical results.

Note that condition (\ref{eq:cond_bals}) is necessary but does not guarantee the uniqueness of the BALS estimates. Sufficient conditions can be derived by studying the rank properties of $\mathbf{M}_1 = \mathbf{S} \diamond \mathbf{X}\mathbf{H}^{\textrm{T}} \in \mathbb{C}^{KT \times N}$ and $\mathbf{M}_2= \mathbf{S} \diamond \mathbf{G} \in \mathbb{C}^{KL \times N}$. To this end, let us invoke the following result.

\

\vspace{-2ex}
\noindent \textit{Lemma 1 (Rank of the Khatri-Rao product \cite{SidNway2000,STEGEMAN2007540})}: For $\mathbf{A} \in \mathbb{C}^{I \times N}$ and $\mathbf{B} \in \mathbb{C}^{J \times N}$, if $\textrm{rank}(\mathbf{A}) \geq 1$ and $\textrm{rank}(\mathbf{B}) \geq 1$, then
$\textrm{rank}(\mathbf{A} \diamond \mathbf{B})\geq \textrm{min}\left(\textrm{rank}(\mathbf{A}) + \textrm{rank}(\mathbf{B}) -1,N\right).$ 
A concise proof of this lemma can be found in \cite{STEGEMAN2007540,LDL2008}. This result means that the Khatri-Rao product of $\mathbf{A}$ and $\mathbf{B}$ will have full column-rank if $\textrm{rank}(\mathbf{A}) + \textrm{rank}(\mathbf{B}) \geq N + 1$.

The application of Lemma 1 to the Khatri-Rao structured matrices $\mathbf{M}_1 = \mathbf{S} \diamond \mathbf{X}\mathbf{H}^{\textrm{T}}$ and $\mathbf{M}_2= \mathbf{S} \diamond \mathbf{G}$ leads to the following conditions that guarantee the uniqueness of the channel estimates via solving the problems (\ref{func costG}) and (\ref{Func costZ})
\begin{eqnarray}
&&\textrm{rank}(\mathbf{S}) + \textrm{rank}(\mathbf{X}\mathbf{H}^{\textrm{T}}) \geq N+1\label{eq:cond1kr}\\
&&\textrm{rank}(\mathbf{S}) + \textrm{rank}(\mathbf{G}) \geq N+1\label{eq:cond2kr}
\end{eqnarray}
Let us consider that our channel training design parameters, namely, the IRS phase shift matrix $\mathbf{S}$ and the pilot symbols matrix $\mathbf{X}$ are designed to have full rank. The above conditions yield useful corollaries for the system design, when BALS is used. In the following, we discuss these corollaries.

\vspace{-1ex}
\subsection{The BS-IRS and IRS-UT channel matrices have full rank}
{\color{black}Assuming that both channel matrices $\mathbf{H}$ and $\mathbf{G}$ have full rank (e.g. i.i.d. Rayleigh fading),
conditions  (\ref{eq:cond1kr})-(\ref{eq:cond2kr}) can be rewritten as
\begin{eqnarray}
&& \textrm{min}(K,N)+ \textrm{min}(M,N) \geq N+1 \label{eq:cond1_simplified}
\\
&& \textrm{min}(K,N)+ \textrm{min}(L,N) \geq N+1.
\label{eq:cond2_simplified}
\end{eqnarray}
}
We may distinguish two cases, as follows.
\begin{itemize}
\item $N\geq T \geq M$ and $N\geq L$: In this scenario, the BS and the UT have small to moderate antenna array sizes, whose number of antennas are smaller than the number of IRS elements. In this case, conditions (\ref{eq:cond1kr})-(\ref{eq:cond2kr}) reduce to
\begin{eqnarray}
&& M + \textrm{min}(K,N)  \geq N+1\label{eq:cond1krv2}\\
&& L + \textrm{min}(K,N)  \geq N+1\label{eq:cond2krv2}
\end{eqnarray}
\item $T \geq M \geq N$: In this scenario, the BS is assumed to be equipped with a large antenna array, which has as many antennas as the number of IRS elements (massive MIMO setup). Since condition (\ref{eq:cond1kr}) is always satisfied regardless of the value of $K$, the uniqueness of the channel estimates only depends on (\ref{eq:cond2kr}), which translates to
\begin{equation}
\textrm{min}(K,N) +  \textrm{min}(L,N) \geq N + 1 \label{eq:cond2krv3}
\end{equation}
\end{itemize}

The conditions (\ref{eq:cond1krv2}) and (\ref{eq:cond2krv2}) establish a tradeoff between the time dimension (number $K$ of IRS training blocks) and the two spatial dimensions (number $M$ and $L$ of transmit and receive antennas, respectively) from a channel recovery viewpoint. For instance, if $K < N$, these conditions imply $M+K \geq N+1$ and $L+K\geq N+1$, which is equivalent to $\min(M+K,L+K) \geq N+1$. Hence, reducing the number of transmit (or receive) antennas should be compensated by an increase on the number of time blocks in order to ensure the uniqueness of the channel estimates via the BALS algorithm.

\vspace{-1ex}
\subsection{The BS-IRS and IRS-UT channel matrices are rank-deficient}\label{subsec2}
In millimeter-wave MIMO systems, the large number of transmit/receive antennas combined with scattering-poor propagation may result in low rank channel matrices $\mathbf{H}$ and $\mathbf{G}$. Let us assume that the signal propagating between the BS and IRS experiences $R_1$ clusters, while the signal propagating between the IRS and the UT experiences $R_2$ clusters. Moreover, assume that each cluster contributes with one ray that has a complex amplitude and an associated angle of arrival/departure. {\color{black}We can represent the BS-IRS and IRS-UT channels as follows \cite{Heath2016}
 \begin{equation}
     \mathbf{H} = \mathbf{A}_{\textrm{IRS}}\textrm{diag}(\boldsymbol{\alpha})\mathbf{A}_{\textrm{BS}}^{\textrm{H}},\label{eq:Hrankdef}
 \end{equation}
  \begin{equation}
     \mathbf{G} = \mathbf{B}_{\textrm{UT}}\textrm{diag}(\boldsymbol{\beta})\mathbf{B}_{\textrm{IRS}}^{\textrm{H}},\label{eq:Grankdef}
 \end{equation}
where $ \mathbf{A}_{\textrm{BS}} \in \mathbb{C}^{M \times R_1}$, $\mathbf{A}_{\textrm{IRS}} \in \mathbb{C}^{N \times R_1}$, $\mathbf{B}_{\textrm{UT}} \in \mathbb{C}^{L \times R_2}$ and $\mathbf{B}_{\textrm{IRS}} \in \mathbb{C}^{N \times R_2}$ are the array response matrices, and the vectors $\boldsymbol{\alpha}$ and $\boldsymbol{\beta}$ hold the complex amplitude coefficients of the BS-IRS and IRS-UT channels, respectively.}  More specifically, we have $\textrm{rank}(\mathbf{H})=R_1$ and $\textrm{rank}(\mathbf{G})=R_2$, where it is assumed that $R_1 \leq \textrm{min}(M,N)$ and $R_2 \leq \textrm{min}(L,N)$ (rank-deficient case).

{\color{black}First, note that the conditions (\ref{eq:cond_krf}) required by the KRF algorithm to solve the decoupled channel estimation problem are not affected by the rank deficiency of the channel matrices. However, this is not the case for BALS, since the uniqueness of the LS estimates of $\mathbf{G}$ and $\mathbf{H}$ depend on the rank of these matrices, as shown in conditions (\ref{eq:cond1kr}) and (\ref{eq:cond2kr}).} Considering BALS, we can draw useful corollaries as follows.
\begin{itemize}
\item $T \geq M$: Conditions (\ref{eq:cond1kr}) and (\ref{eq:cond2kr}) reduce to
\begin{eqnarray}
&& \textrm{min}(K,N) + R_1 \geq N+1 \label{eq:cond2krv4} \\
&& \textrm{min}(K,N) + R_2 \geq N+1 \label{eq:cond2krv4_2}
\end{eqnarray}
It is worth discussing the following cases. If $K\geq N$, we conclude that these conditions are always satisfied, irrespective of the ranks of the channel matrices. If $K<N$, the these conditions reduce to $K+R_1\geq N+1$ and $K+R_2\geq N+1$, yielding a useful design recommendation the number $K$ of  blocks that guarantee the uniqueness of the channel estimates in the rank-deficient case.
\item $K \geq N$: In this case, conditions (\ref{eq:cond1kr}) and (\ref{eq:cond2kr}) are always satisfied, irrespective of the rank of  $\mathbf{G}$ and $\mathbf{H}$.
\end{itemize}

\noindent \textit{Discussion}: It is worth noting that the proposed channel estimation methods still work for $K=1$. However, in this setup only the cascaded channel $\mathbf{C}=\mathbf{G}\textrm{diag}(\mathbf{s})\mathbf{H}$ can be estimated. The performance enhancements that come from the decoupling of the estimates of $\mathbf{H}$ and $\mathbf{G}$ (via KRF or BALS) cannot be obtained. Otherwise stated, leveraging extra training time diversity by increasing the number $K$ of IRS phase shift patterns allows us to extract additional gains in comparison to the traditional LS estimator, as will be shown in our numerical results. However, such gains come at the expense of an increase on the training resources. Therefore, here we clearly see a trade-off between training overhead and performance.

In addition, it is clear from conditions (\ref{eq:cond1kr})-(\ref{eq:cond2kr}), or equivalently, from conditions (\ref{eq:cond1_simplified}) and (\ref{eq:cond1_simplified}), that BALS can operate under more flexible choices for $K$ than KRF, since the latter always requires $K\geq N$. Note that, for $M\geq N$ and $L\geq N$, these conditions are satisfied even for small values of $K$. In practice, this means that BALS may operate with a much lower training overhead than KRF and the traditional LS methods. However, our experience shows that for a large number of IRS elements, working with small values of $K$ results in a slower convergence speed of BALS due to the limited time diversity. Therefore, a trade-off between training overhead and complexity arises when BALS is considered.
		
Note also that, in terms of the required training time resources, BALS becomes equivalent to KRF in the single transmit antenna case ($M=1$) and/or in the single receive antenna case ($L= 1$). Otherwise stated, for MISO and/or SIMO IRS-assisted systems, BALS and KRF have exactly the same requirement $K\geq N$. Thus, we can say that BALS is advantageous over KRF in terms of training overhead when considering the MIMO case. Likewise, performance gains of KRF and BALS over the baseline LS method also arise in the MIMO setup, where spatial degrees of freedom at the transmitter and the receiver are efficiently exploited to obtain more accurate channel estimates. {\color{black} Finally, one can note that the channel estimates $\hat{\mathbf{G}}$ and $\hat{\mathbf{H}}$ are affected by scaling factors$\footnote{The permutation ambiguity inherent to blind estimation is not present due to the knowledge of the IRS phase shift matrix $\mathbf{S}$ at the receiver.}$ satisfying $\hat{\mathbf{H}}= \boldsymbol{\Delta}_H\mathbf{H}$ and $\hat{\mathbf{G}}  = \mathbf{G}\boldsymbol{\Delta}_G$, where $\boldsymbol{\Delta}_H\boldsymbol{\Delta}_G=\mathbf{I}_N$. These scaling ambiguities are irrelevant in our context since they compensate each other when building the estimate of the cascade BS-IRS-UT channel.}
		
\section{Generalizations to multi-user scenarios}
Although we have focused on the single BS and single UT scenario, the proposed approach as well as the derived results can be easily generalized and adapted to IRS-assisted  muliple-access/multi-user MIMO systems. Let us take the uplink case as an example. The downlink case follows exactly the same model by just inverting the roles of BS and UT. We can distinguish two scenarios, which are discussed as follows.

\subsection{Multiple users communicate with a single BS via the IRS}
Let us consider $U$ UTs communicating with a single BS via the IRS. The direct link between the UTs and the BS is assumed to be too weak or unavailable. Assuming for simplicity that all the users have the same number $L$ of transmit antennas, we can adapt equation (\ref{eq:PARAFAC MODEL2}) such that the contribution of the $u$-th user to the received signal at the BS  is given as
\begin{equation}
\begin{aligned}
\overline{\mathbf{Y}}_u[k]& = \mathbf{H}^{\textrm{T}}\mathbf{D}_k(\mathbf{S})\mathbf{G}_u^{\textrm{T}}\mathbf{X}_u^{\textrm{T}}
\end{aligned}
\end{equation}
where $\mathbf{X}_u \in \mathbb{C}^{T \times L}$ and $\mathbf{G}_u \in \mathbb{C}^{L \times N}$ are respectively the $u$-th user pilot matrix and uplink channel matrix. Note that the IRS-BS channel $\mathbf{H}$ is common to all the users. The total signal received from the $U$ users at the $k$-th time block, in the noiseless case, can then be expressed as
\begin{equation}
\begin{aligned}
\overline{\mathbf{Y}}[k]& = \mathbf{H}^{\textrm{T}}\mathbf{D}_k(\mathbf{S})\left(\mathbf{X}_1\mathbf{G}_1\right)^{\textrm{T}} + \dots + \mathbf{H}^{\textrm{T}}\mathbf{D}_k(\mathbf{S})\left(\mathbf{X}_U\mathbf{G}_U\right)^{\textrm{T}}\\
&=\mathbf{H}^{\textrm{T}}\mathbf{D}_k(\mathbf{S})\left[\sum\limits_{u=1}^U(\mathbf{X}_u\mathbf{G}_u)^{\textrm{T}}\right].
\end{aligned}
\label{eq:parafac_multi-user}
\end{equation}
Defining $\overline{\mathbf{X}}\doteq [\mathbf{X}_1, \ldots, \mathbf{X}_U] \in \mathbb{C}^{T \times UL}$, and $\overline{\mathbf{G}}\doteq [\mathbf{G}^{\textrm{T}}_1, \ldots, \mathbf{G}^{\textrm{T}}_U]^{\textrm{T}} \in \mathbb{C}^{UL \times N}$,  equation (\ref{eq:parafac_multi-user}) translates to\footnote{Note that the positions of $\mathbf{H}$ and $\mathbf{G}$ in are swapped in (\ref{eq:PARAFAC_MULT_USER__ONE_BS}) compared to (\ref{eq:PARAFAC MODEL2}) in addition to transposition, since channel reciprocity is assumed for the UT-IRS and IRS-BS links.}
\begin{equation}
\mathbf{Y}[k] =     \mathbf{H}^{\textrm{T}}\mathbf{D}_k(\mathbf{S})\overline{\mathbf{Z}}^{\textrm{T}} 
\quad \overline{\mathbf{Z}}\doteq \overline{\mathbf{X}}\overline{\mathbf{G}} \in \mathbb{C}^{T \times N}.
\label{eq:PARAFAC_MULT_USER__ONE_BS}
\end{equation}
Comparing (\ref{eq:PARAFAC_MULT_USER__ONE_BS}) with (\ref{eq:PARAFAC MODEL2}), we can see that the multi-user signal model has the same tensor structure as the single-user one, the essential difference being on the definition of the factor matrix $\overline{\mathbf{Z}}$ which is now given by inner product of block matrices composed of $U$ blocks (each having $L$ columns as in the single-user scenario). Otherwise stated (\ref{eq:PARAFAC_MULT_USER__ONE_BS}) corresponds to a PARAFAC model of $\overline{\mathcal{Y}} \in \mathbb{C}^{M \times T \times K}$ with factor matrices $(\mathbf{H}^{\textrm{T}},\overline{\mathbf{X}}\overline{\mathbf{G}}, \mathbf{S})$, {\color{black} and unfoldings $1$-mode and $2$-mode given as $\overline{\mathbf{Y}}_{1} = \mathbf{H}^{\textrm{T}}(\mathbf{S} \diamond \overline{\mathbf{X}}\overline{\mathbf{G}})^{\textrm{T}}$ and $\overline{\mathbf{Y}}_{2} = \overline{\mathbf{X}}\overline{\mathbf{G}}(\mathbf{S} \diamond \mathbf{H}^{\textrm{T}})^{\textrm{T}}$, respectively.} Since the structure of the tensor model is not changed, both KRF and BALS algorithms can be directly applied to the multi-user model (\ref{eq:PARAFAC_MULT_USER__ONE_BS}) under more restrictive choices for $T$, {\color{black}due to the fact the full rankness of $\mathbf{X}$ now requires $T\geq UL$. In this case, assuming that the channel matrices have full rank, the application of Lemma 1 leads to
\begin{eqnarray}
&& \textrm{min}(K,N)+ \textrm{min}(UL,N) \geq N+1 \label{eq:cond1krv2_MULTIUSER2} \\
&& \textrm{min}(K,N)+ \textrm{min}(M,N) \geq N+1.
\label{eq:cond2krv2_MULIUSER2}
\end{eqnarray}
}
These conditions are analogous to  (\ref{eq:cond1_simplified})-(\ref{eq:cond2_simplified}), by exchanging the roles of $M$ and $L$, while adding the factor $U$.

\subsection{Multiple users communicate with multiple BSs via the IRS}
We consider that $P$ BSs receive the signals transmitted by the $U$ users via the IRS. Without loss of generality, the BSs are assumed to be equipped with the same number $M$ of antennas. The model (\ref{eq:PARAFAC_MULT_USER__ONE_BS}) is only slightly modified by adding a dependency of the received signal on the index $p$ of the receiving BS, i.e.,
\begin{equation}
\overline{\mathbf{Y}}_p[k]=\mathbf{H}^{\textrm{T}}_p\mathbf{D}_k(\mathbf{S})\overline{\mathbf{Z}}^\textrm{T},\quad \overline{\mathbf{Z}}\doteq \overline{\mathbf{X}}\overline{\mathbf{G}}. \label{eq:parafac_multi-user_multiBS}	
\end{equation}
In particular, in a cooperative setting where the $P$ BSs communicate (e.g. via a common backhauling structure), we can derive an equivalent augmented signal model as follows
\begin{equation}
\overline{\overline{\mathbf{Y}}}[k]= \left[\begin{array}{c} \hspace{-1ex}
\overline{\mathbf{Y}}_1[k] \hspace{-1ex}\\ \hspace{-1ex}\vdots\hspace{-1ex} \\ \hspace{-1ex}\overline{\mathbf{Y}}_P[k]\hspace{-1ex}\end{array}\right]= \left[\begin{array}{c}
\mathbf{H}^{\textrm{T}}_1\\ \vdots \\ \mathbf{H}^{\textrm{T}}_P\end{array}\right]
\mathbf{D}_k(\mathbf{S})\overline{\mathbf{Z}}^\textrm{T}=\overline{\mathbf{H}}^{\textrm{T}}\mathbf{D}_k(\mathbf{S})\overline{\mathbf{Z}}^\textrm{T}, \label{eq:parafac_multi-user_multiBS2}	
\end{equation}
where $\overline{\mathbf{H}}\doteq [\mathbf{H}_1, \ldots, \mathbf{H}_P]^{\textrm{T}} \in \mathbb{C}^{PM \times N}$ is the composite channel combining the IRS links to the $P$ BSs. The received signal (\ref{eq:parafac_multi-user_multiBS2}) corresponds to a PARAFAC model of $\overline{\overline{\mathcal{Y}}} \in \mathbb{C}^{PM \times T \times K}$ with factor matrices
$(\overline{\mathbf{H}},\overline{\mathbf{X}}\overline{\mathbf{G}}, \mathbf{S})$. Note that, differently from the single-user single-BS model (\ref{eq:PARAFAC MODEL2}) and the multi-user single-BS model (\ref{eq:PARAFAC_MULT_USER__ONE_BS}), in the multi-user multi-BS model (\ref{eq:parafac_multi-user_multiBS2}) the dimensionality of the first mode of the received signal tensor has been increased by a factor $P$ due to the assumption of cooperating BSs. In this scenario, condition (\ref{eq:cond1krv2_MULTIUSER2}) remains the same, while condition (\ref{eq:cond2krv2_MULIUSER2}) slightly changes to
$\textrm{min}(K,N)+ \textrm{min}(PM,N) \geq N+1$.

As a final remark, in terms of receiver processing, it is clear that both KRF and BALS have an increased computational complexity in the discussed multi-user scenarios, due to the increased dimensionality of the channel matrices $\overline{\mathbf{G}}$ and $\overline{\mathbf{H}}$.

\section{Numerical Results}
In this section, several numerical results are presented to evaluate the performance of the proposed channel estimation methods, while comparing to competing approaches. We also evaluate the CRB as a reference for comparisons. The channel estimation accuracy is evaluated in terms of the normalized mean square error (NMSE) given by
\begin{equation}
    \textrm{NMSE}(\hat{\mathbf{H}}) = \textcolor{black}{\frac{1}{R}\sum_{r=1}^{R} \dfrac{\|\mathbf{H}^{(r)} - \hat{\mathbf{H}}^{(r)}\|_F^2}{ \|\mathbf{H}^{(r)}\|_F^2}},
\end{equation}
where $\hat{\mathbf{H}}^{(r)}$ is the BS-IRS channel estimated at the $r$-th run, and $R$ denotes the number of Monte Carlo runs. The same definition applies to the estimated IRS-UT channel.
The SNR (in dB) is defined as
\begin{equation}
    \textrm{SNR} = 10\textrm{log}_{10}(\|[\overline{\mathcal{Y}}]\|_F^2/ \|[\mathcal{B}]\|_F^2),
\end{equation}
where $\overline{\mathcal{Y}}$ is the noiseless received signal tensor generated according to (\ref{scalar_Y}), and $\mathcal{B}$ is the additive noise tensor.
\begin{figure}[!t]
\begin{center}
	\includegraphics[scale=0.57]{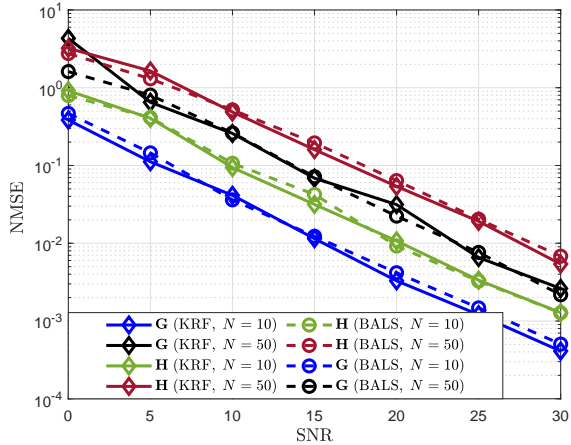}
\end{center}
	\vspace{-3ex}
\caption{NMSE of the estimated channels $\hat{\mathbf{H}}$ and $\hat{\mathbf{G}}$.}
	\label{COMPGH}
\end{figure}
\begin{figure}[!t]
\begin{center}
	\includegraphics[scale=0.57]{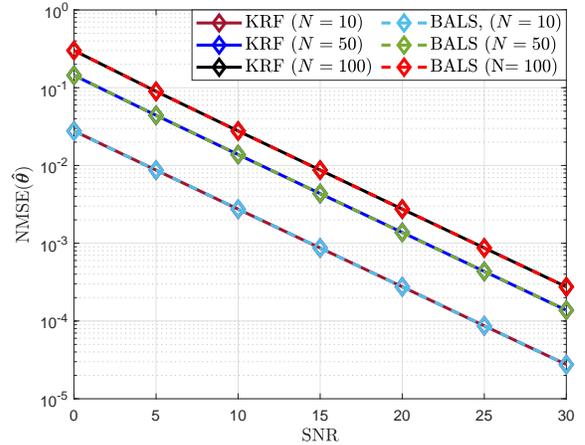}
\end{center}
	\vspace{-3ex}
\caption{NMSE for the equivalent channel $\hat{\boldsymbol{\theta}}$.}
	\label{COMPHC}
\end{figure}

\begin{figure}[!t]
\begin{center}
	\includegraphics[scale=0.57]{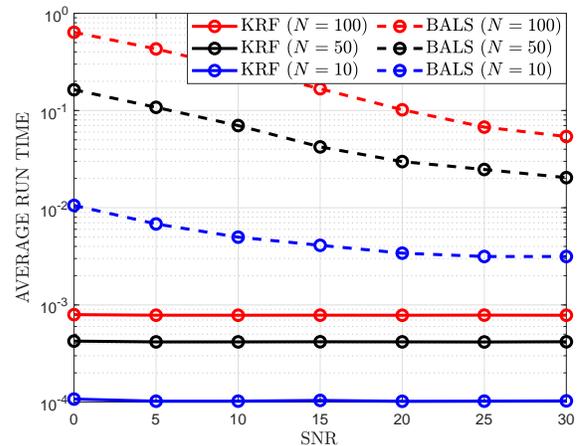}
\end{center}
	\vspace{-3ex}
\caption{Average runtime of KRF and BALS algorithms.}
	\label{RTCOMP}
\end{figure}

\begin{figure}[!t]
\begin{center}
  \includegraphics[scale=0.57]{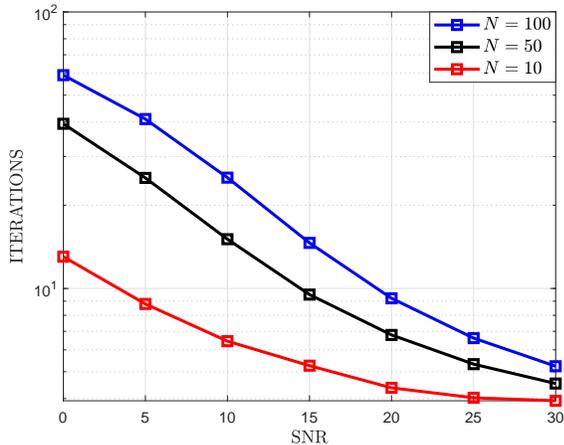}
\end{center}
    \vspace{-3ex}
    \caption{Number of iterations to convergence of the BALS algorithm.}
    \label{COMPIT}
\end{figure}

We assume that the entries of the BS-IRS and IRS-UT channel matrices $\mathbf{H}$ and $\mathbf{G}$ are independent and identically distributed zero-mean circularly-symmetric complex Gaussian random variables. The Figures \ref{COMPGH}, \ref{RTCOMP} and \ref{COMPIT}, represent an average from $R=5000$ run Monte Carlo runs for the fixed system parameters $\{T = 4, L = 2, K = 50, M = 3\}$ and $N \in \{50, 100\}$.

Figure \ref{COMPGH} depicts the NMSE vs. SNR curves for the KRF and the BALS algorithms. We can see that both algorithms provide satisfactory performances. The performance degrades as the number of IRS elements is increased, which is an expected result since the number of channel coefficients in $\mathbf{G}$ and $\mathbf{H}$ to be estimated also increases with $N$. {\color{black} In Figure \ref{COMPHC}, the NMSE of the composite parameter vector $\boldsymbol{\theta}$ is shown. The parameters used get this figure was $K = 100$, $M = 3$, $T = 4$, $L = 20$, $N \in \{10, 50, 100\}$ and $1000$ Monte Carlo runs. The results are in line with those of the previous figure, where we observe a performance degradation as $N$ is increased, which confirms our expectations. {\color{black} An approach to overcome such a performance degradation is to partition the IRS into groups, and activate/deactivate each group in a sequential way in the time domain, so that the at each time, the sub-channels associated with a given group are estimated \cite{C_You_2019,R_Zhang_Blocks_Channel,R_Zhang_Review2}. This approach, however, would increase the total training time by a factor corresponding to the number of groups.}
}
The following experiments compare the average runtime of KRF and BALS. The results are depicted in Figure \ref{RTCOMP}, and corroborate the higher complexity of BALS compared to KRF. Note that the runtime of BALS grows faster then that of KRF with the increase on the number $N$ of IRS elements. On the other hand, as we pointed out earlier (comparison between (\ref{eq:cond_krf}) and (\ref{eq:cond_bals})), BALS can operate under less restrictive choices (smaller values) for $K$ in comparison to KRF. Hence, there is a tradeoff between complexity and operating conditions for the two proposed channel estimation methods.

In Figure \ref{COMPIT}, we evaluate the required number of iterations of the BALS algorithm to achieve the convergence according to the criterion discussed in Section \ref{sec:bals_parafac}. We can note that the required number of iterations grows with $N$, as expected. The difference in the convergence speed for different values of $N$ is more pronounced in the low SNR range. For high SNRs, the convergence becomes less sensitive to $N$.

\begin{figure}[!t]
\begin{center}
	\includegraphics[scale=0.57]{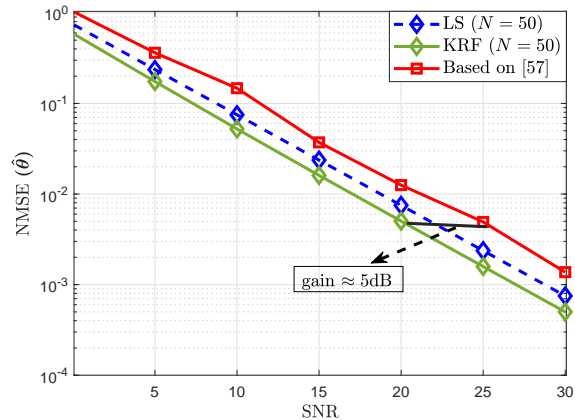}
\end{center}
	\vspace{-3ex}
\caption{NMSE of the estimated cascaded channel via the method of \cite{Laing_Liu2020}.}
	\label{COMPARe with LIANG}
\end{figure}

In Figure \ref{COMPARe with LIANG}, we consider the uplink multi-BS scenario, which follows the signal model (\ref{eq:parafac_multi-user_multiBS2}). We assume $P=2$, $M=1$, and $U=1$. We compare the KRF receiver with a competing channel estimation method proposed recently in \cite{Laing_Liu2020}, which considers the single-antenna multi-user reception scenario\footnote{In \cite{Laing_Liu2020} the authors assume multiple receiving single-antenna UTs and a single multi-antenna BS in the downlink, while our model (\ref{eq:parafac_multi-user_multiBS2}) assumes multiple receiving BSs and a single multi-antenna UT in the uplink. Due to the channel reciprocity assumption, the signal model of \cite{Laing_Liu2020} is equivalent to our signal model (\ref{eq:parafac_multi-user_multiBS2}).  For a fair comparison, we assume $P=2$ UTs for the channel estimation method of \cite{Laing_Liu2020}.  In this case, the dimensions of the channel matrices are exactly the same for both methods.}.
Therein, the channel estimation requires three sequential stages, i.e., three time windows. In the first stage, the direct channel is estimated. In the second one, the equivalent channel between the first user and the BS is estimated. Finally, in the third stage, the channel associated with the remaining users are estimated. Similar to our model, in \cite{Laing_Liu2020} the equivalent channel is obtained by stacking the contributions of the $U$ users, i.e., $\mathbf{G} \diamond \mathbf{H}=\big[\left( \mathbf{H}\textrm{diag}(\mathbf{g}_{1})\right)^{\textrm{T}}, \ldots, (\mathbf{H}\textrm{diag}(\mathbf{g}_{U}))^{\textrm{T}}\big]^{\textrm{T}}$.
We can see that  KRF outperforms the competing method, providing an SNR gain of nearly $5$dB. Indeed, KRF jointly estimates all the involved channels in a single training stage, while in \cite{Laing_Liu2020} the channel estimation is carried out in a sequential way, which can induce error propagation. This is a key difference that explains the performance gap in Figure \ref{COMPARe with LIANG}.
\begin{figure}[!t]
\begin{center}
	\includegraphics[scale=0.57]{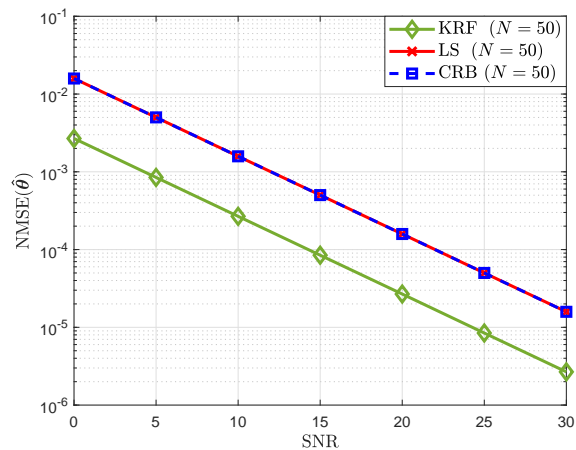}
\end{center}
	\vspace{-3ex}
\caption{Normalized CRB for the equivalent Khatri-Rao channel $\boldsymbol{\theta}$.}
	\label{CRBCOMP}
\end{figure}

In Figure \ref{CRBCOMP}, we compare the results of the proposed KRF method with the conventional LS method. {\color{black}In this experiment, we consider $K=N= 50$, $T=M= 20$, $L= 8$, and $1000$ Monte Carlo runs}. The CRB derived in Appendix A (equations (\ref{eq:crb1})- (\ref{eq:crb1})) is also plotted here as a reference for comparison. Recall that the CRB considers the equivalent linear model obtained from the vectorized version of the received signal model given in (\ref{Y3final}), which we repeat here for convenience
$$
\mathbf{y} = \textrm{vec}\left(\mathcal{Y}\right)= \mathbf{U}\boldsymbol{\theta} +  \mathbf{b},
$$
where $\mathbf{U} = \left(\mathbf{S} \otimes \tilde{\mathbf{X}}\right)$ and $\boldsymbol{\theta} = \textrm{vec}\left(\mathbf{H}^{\textrm{T}} \diamond \mathbf{G}\right)  \in \mathbb{C}^{MNL}$ is the parameter vector consisting of  a vectorized version of the (Khatri-Rao structured) channel matrix combining the IRS-UT and the BS-IRS channel matrices. The conventional LS method plotted in the figure estimates this vectorized channel parameter as $\hat{\boldsymbol{\theta}}= \mathbf{U}^{\dagger}\mathbf{y}$, which ignores the Khatri-Rao structure that is lost in the vectorization of the signal model. In contrast, the proposed KRF method exploits the Khatri-Rao channel structure, and builds $\hat{\boldsymbol{\theta}}$ from the decoupled estimates $\hat{\mathbf{H}}$ and $\hat{\mathbf{G}}$ obtained according to Algorithm 1. 

\begin{figure}[!t]
\begin{center}
	\includegraphics[scale=0.57]{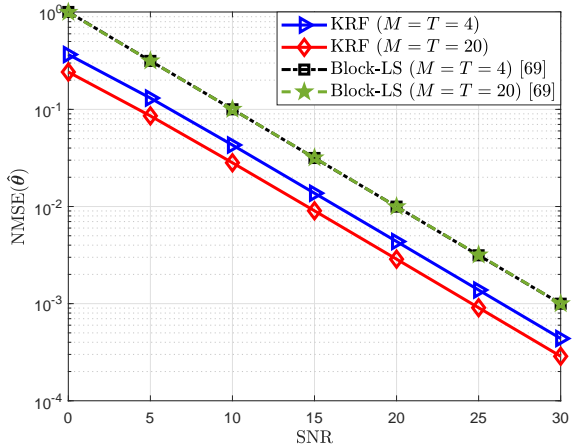}
\end{center}
	\vspace{-3ex}
\caption{NMSE of the $\boldsymbol{\hat{\theta}}$ assuming that  $\hat{\mathbf{H}}$ and $\hat{\mathbf{G}}$ are rank-deficient.}
	\label{COMP_Hc_iid_LR}
\end{figure}
\begin{figure}[!t]
\begin{center}
	\includegraphics[scale=0.57]{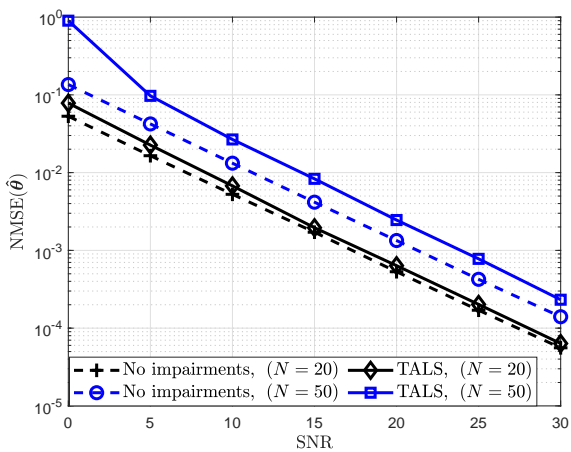}
\end{center}
	\vspace{-3ex}
\caption{TALS performance under IRS amplitude/phase perturbations.}
	\label{Theta with HWI}
\end{figure}
We can see that the LS solution attains the CRB. Furthermore, an interesting result can be noted. The proposed KRF algorithm outperforms the LS solution. The gain in terms of SNR is around $7$dB. This result is explained by the fact that KRF effectively exploits the Khatri-Rao structure that is present in the equivalent channel model. Note that the KRF method solves the problem by reshaping the $ML \times N$ Khatri-Rao channel in the form of $N$ IRS subchannels of dimension $M \times L$, which provides a noise rejection gain thanks to the rank-1 approximation steps. Naturally, when $M$ and $L$ increase (which is the case, for instance, when assuming massive antenna arrays at the BS and UT), the larger is the spreading of the noise across the noise subspace and, consequently, higher levels of noise rejection will be achieved. This is a distinctive feature of the KRF method that is not exploited by the conventional LS channel estimator.

In Figure \ref{COMP_Hc_iid_LR}, we assume that the channel matrices $\mathbf{H}$ and $\mathbf{G}$ are rank-deficient. {\color{black}  In this experiment, the channel matrices are generated according to the model given in (\ref{eq:Hrankdef})-(\ref{eq:Grankdef}). We assume uniform linear arrays at the BS and UT. The IRS has a uniform rectangular array structure. The angles of arrival (AoAs) and angles of departure (AoDs) are randomly generated according to a uniform distribution. At each Monte Carlo run, the azimuth and elevation angles are drawn within the intervals $[-\pi/2, \pi/2]$ and $[0, \pi/2]$, respectively. We consider a single path scenario, where $R_1 = R_2 = 1$, and assume $K=N=64$, $L = 4$ and $T = M \in \{4,20\}$. As a reference for comparison, we also plot the NMSE of the LS-based channel estimation method proposed in \cite{Ref_COMPII}. Therein, the time-domain pilot protocol is the same as the one considered in this work, which consists of dividing the total training time into $K$ blocks across which the phase shift pattern of the IRS is varied. In \cite{Ref_COMPII}, a two-stage scheme is proposed. In the first stage, the cascaded channel $\mathbf{C}_k = \mathbf{G}\mathbf{D}_k(\mathbf{S})\mathbf{H}$ associated with every time block $k$ is individually estimated via an LS method. We refer to this approach as a ``block-LS'' method. The second stage extracts the AoA and AoD parameters by combining the $K$ cascaded channel matrices. Since our method does not estimate the angular parameters of the channel matrices, we compare the proposed KRF method with the first stage of the block-LS method of \cite{Ref_COMPII}, which also provides the unstructured estimate of the channel matrices $\mathbf{H}$ and $\mathbf{G}$. We can see that KRF outperforms block-LS in the two considered system setups. Note that the performance of the block-LS method is not affected when the number $M$ of transmit antennas (assuming $T = M$) is increased. This is in contrast to the KRF method that provides more accurate channel estimates for larger antenna arrays. In particular, the SNR gain of KRF over block-LS is nearly $3.5$ dB for $M=T=4$, and increases to  $5.5$dB for $M=T=20$. Indeed, higher values of $M$ and/or $L$ imply higher levels of noise rejection provided by the KRF method \emph{via} exploiting the Khatri-Rao structure of the cascaded channel. These gains come at the expense of an increased computational complexity, as well as an increase on the length of the pilot sequences.}
{\color{black}Note that the channel ranks $R_1$ and $R_2$ do not need to be known by our channel estimation methods. Nevertheless, a performance enhancement could be obtained by exploiting the knowledge of these ranks (see, e.g. methods like \cite{Wang2020}), or, alternatively, by means of compressed sensing methods that capitalize on sparse representations of the channel matrices $\mathbf{H}$ and $\mathbf{G}$. This is an interesting topic for a future research.}

{\color{black} In Figure \ref{Theta with HWI}, we assume that the IRS is affected by amplitude and phase perturbations, as well as unknown blockages, due to hardware and/or environmental-induced impairments. In this scenario, the receiver has an imperfect knowledge of the IRS phase shift  matrix. The estimation/refinement of these phase shifts is carried out using the TALS algorithm, which extends BALS by including an additional LS estimation step associated with the update of $\mathbf{S}$ within the loop, as discussed in Section III-D. To model these impairments, we assume $s_{k,n} = \left(a_{k,n}f_{k,n}\right)\bar{s}_{k,n}$, where $\bar{s}_{k,n}$ is the originally designed phase shift (i.e., ($k,n$)-th element of a DFT matrix), $a_{k,n} \in \{0, 1\}$ models the presence or not of a blockage at the $n$-th element and $k$-th block, $f_{k,n} \sim CN(0,\gamma)$ models the hardware impairments \cite{Badiu2020,Erwu2020}, and $\gamma$ denotes the variance of these perturbations. In the TALS  algorithm, the initialization of $\hat{\mathbf{S}}$ is chosen as a DFT matrix, following its original design. This initialization always provides better results than a random one. Our simulations assume $T= M= 50$, $K=100$, $L=4$, $\gamma = 0.01$, 20\% blocked IRS elements, and i.i.d. channel matrices. Although a performance degradation is observed in comparison with the perfectly-known IRS phase shifts case, we can see that the TALS algorithm can handle this more challenging scenario. In particular, the NMSE gap with respect to the ideal case increases as more elements are used in the IRS. Indeed, in the imperfect IRS scenario, the total number of parameters to be estimated by the TALS algorithm is $(K+M+L)N$, in contrast to $(M+L)N$ in the perfectly-known IRS case, implying an addition of $KN$ unknown factors.}

\section{Conclusion {\color{black} and Perspectives}}
We have proposed novel pilot-assisted receiver designs for IRS-assisted MIMO communication systems via a tensor modeling approach. The proposed KRF and BALS receivers effectively exploit the tensor structure that is present in the received signal. Both solutions yield decoupled estimates of the BS-IRS and IRS-UT channels at the receiver for a passive IRS. The closed-form KRF method has a lower complexity but a more restrictive requirement on the training parameter $K$, while the iterative BALS method, although being more computationally complex, can operate under more flexible choices for this parameter with a lower training overhead. Our design recommendations provide useful conditions on the system parameters that guarantee the uniqueness of the channel estimates. Our numerical results have demonstrated the superior performance of KRF and BALS compared to the conventional LS estimator, which ignores the Khatri-Rao structure of the combined channel matrix.
{\color{black} In addition, the proposed tensor modeling approach allows to deal with a nonideal setup where the IRS phase shifts are not perfectly known at the receiver due to phase perturbations/fluctuations. In this more difficult setup, leveraging the trilinear structure of the received signal by means of a TALS algorithm provides us a joint estimation of the channel matrices and the IRS phase shift matrix. The proposed solutions also provide better results than recently proposed competing methods. Generalizations of our tensor modeling approach to multi-user scenarios have also been discussed, and analytical expressions for the CRB have been derived. The proposed approach can easily be extended to better deal with the millimeter wave scenario by assuming hybrid analog digital structures at the BS and UT sides. Combining the proposed algorithms with compressed sensing methods could provide further performance enhancements for low-rank (sparse) channels. In addition, leveraging to data-driven receivers capable of a joint channel estimation and symbol recovery would be desirable to save training resources. To this end, an extension of the proposed tensor modeling approach to the semi-blind case is a perspective of this work. }


\begin{appendices}
\section{\textcolor{black}{Expected Cram\'er Rao lower Bound}}
\label{Appendix A}

\textcolor{black}{In the following, we derive the closed-form CRB  expressions for the channel estimation problem proposed in this work. The CRB provides the lower bound on the variance of achieved by an unbiased estimator. 
If $\boldsymbol{\hat{\theta}}$ is an unbiased estimate of $\boldsymbol{\theta}$, the NMSE measurements is lower bounded by the CRB such as,
\begin{equation}
    \mathbb{E}\|\boldsymbol{\theta} - \boldsymbol{\hat{\theta}}\|^2 \geq Tr\{\textrm{CRB}{\boldsymbol{(\theta)}}\},
\end{equation}
where CRB($\boldsymbol{\theta}$) is given as the inverse of the Fisher Information Matrix (FIM), denoted by $\mathbf{F}(\boldsymbol{\theta})$, such as
\begin{equation}
    \textrm{CRB}(\boldsymbol{\theta}) \geq \mathbf{F}(\boldsymbol{\theta})^{-1}.
\end{equation}}

An extension for \textcolor{black}{complex-valued parameters} is derived in \cite{Favier2019} by working on the structured parameter vector $\boldsymbol{\theta}_c = \left[ \boldsymbol{\bar{\theta}}^{\textrm{T}} \boldsymbol{\widetilde{\theta}}^{\textrm{T}} \right]^{\textrm{T}}$, where $\boldsymbol{\bar{\theta}}=Re(\boldsymbol{\theta})$, and $\boldsymbol{\widetilde{\theta}}=Im(\boldsymbol{\theta})$. Thereby, with a nuisance parameter $\gamma$, the CRB for \textcolor{black}{complex-valued random parameters} is given as
\begin{equation}\label{CRB_tot}
    \mathbb{E}\|\boldsymbol{\theta}_c - \boldsymbol{\hat{\theta}}_c\|^2 \geq \mathbb{E}_{\boldsymbol{\bar{\theta}},\boldsymbol{\widetilde{\theta}},\gamma}\left\{Tr\{\textrm{CRB}\boldsymbol{(\bar{\theta})}\} + Tr\{\textrm{CRB}\boldsymbol{(\widetilde{\theta})}\}\right\}.
\end{equation}

\textcolor{black}{
For an observation vector that follows a complex circular Gaussian distribution, $\mathbf{y} \sim CN(\mu,\mathbf{R})$, a useful way used to obtain the FIM is to use the Slepian-Bangs (SB) formula \cite{P_Stoica}:
\begin{eqnarray}
\left[ \mathbf{F}(\boldsymbol{\theta}) \right]_{i,j} &=& 2Re\left\{ \left( \dfrac{\partial \mu}{\partial\boldsymbol{[\theta}]_i}\right)^{H} \mathbf{R}^{-1}\left( \dfrac{\partial \mu}{\partial\boldsymbol{[\theta}]_j}\right) \right \}
\label{SB_Formula_PART1}\\
& + & Tr \left \{\left( \dfrac{\partial \mathbf{R}}{\partial\boldsymbol{[\theta}]_i}\right)\mathbf{R}^{-1} \left( \dfrac{\partial \mathbf{R}}{\partial\boldsymbol{[\theta}]_j}\right)\mathbf{R}^{-1} \right \}.
\label{SB_Formula_PART2}
\end{eqnarray}
}
%
%
Let us recall (\ref{unfolding3}):
\begin{equation}
\begin{aligned}
\left[\mathcal{Y}\right]_{(3)} &=& \mathbf{S}\left[\left( \mathbf{X} \otimes \mathbf{I}_L\right)\left(\mathbf{H}^{\textrm{T}} \diamond \mathbf{G}\right) \right]^{\textrm{T}}\\
&=&\mathbf{S}\left(\mathbf{H}^{\textrm{T}} \diamond \mathbf{G}\right)^{\textrm{T}}\left( \mathbf{X} \otimes \mathbf{I}_L\right)^{\textrm{T}},
\end{aligned}
\label{CRB}
\end{equation}
or, equivalently,
\begin{equation}
  \left[\mathcal{Y}\right]_{(3)}^{\textrm{T}} = \left( \mathbf{X} \otimes \mathbf{I}_L \right) \left( \mathbf{H}^{\textrm{T}} \diamond \mathbf{G} \right) \mathbf{S}^{\textrm{T}}.
  \label{DERIVECRBFROMHERE}
\end{equation}

Considering the vectorized version of the 3-mode unfolding $\left[\mathcal{Y}\right]_{(3)}^{\textrm{T}}$, the following linear model with respect to the parameters of interest is obtained according to
\begin{equation}\label{Linear_MODEL CRB}
\mathbf{y} = \textrm{vec}\left(\left[\mathcal{Y}\right]_{(3)}^{\textrm{T}}\right)= \mathbf{U}\boldsymbol{\theta},
\end{equation}
where $\mathbf{U} = \left( \mathbf{S} \otimes \mathbf{X} \otimes \mathbf{I}_L \right)$, and
\begin{equation}
\boldsymbol{\theta} = \textrm{vec}\left(\mathbf{H}^{\textrm{T}} \diamond \mathbf{G}\right)  \in \mathbb{C}^{MNL}\label{eq:vectheta}
\end{equation}
denotes the vectorized version of the Khatri-Rao structured channel.
From the observation vector $\mathbf{y}$ given by (\ref{Linear_MODEL CRB}), the statistics of the noisy  observation is given by
\begin{equation}
    \mathbf{y} \sim CN\left( \boldsymbol{\mu}_1, \mathbf{R}_1 \right),
\end{equation}
where,
\begin{eqnarray}
\boldsymbol{\mu}_1 &=& \mathbf{U}\boldsymbol{\theta},\\
\mathbf{R}_1 & = & \sigma^2\mathbf{I}.
\end{eqnarray}

\textcolor{black}{
As $\mathbf{R}_1$ parameter-invariant, the second term of the SB formula vanishes, hence the $(2MNL) \times (2MNL)$ FIM, obtained after the calculation from (\ref{SB_Formula_PART1}), is given by
\begin{equation}
    \mathbf{F}(\boldsymbol{\theta}_c) = \dfrac{2}{\sigma^2}\left[
     \begin{array}{lr}
          Re\{\mathbf{U^{H}}\mathbf{U}\} &  -Im\{\mathbf{U^{H}}\mathbf{U}\}\\
           Im\{\mathbf{U^{H}}\mathbf{U}\}^{\textrm{T}}& Re\{\mathbf{U^{H}}\mathbf{U}\}
     \end{array}
    \right].
\end{equation}}

\textcolor{black}{Considering the trace and the inverse of a $2\times 2$ block matrix, we obtain}
\textcolor{black}{
\begin{equation}
Tr\{\textrm{CRB}(\bar{\boldsymbol{\theta}})\} = \dfrac{\sigma^2}{2}Tr\left\{ \left( \bar{\mathbf{M}} + \widetilde{\mathbf{M}}\bar{\mathbf{M}}^{-1}\widetilde{\mathbf{M}} \right)^{-1} \right\},
\label{CRB_Real_PART}
\end{equation}
\begin{equation}
\begin{split}
Tr\{\textrm{CRB}(\widetilde{\boldsymbol{\theta}})\}  =   \dfrac{\sigma^2}{2} & Tr \left\{\bar{\mathbf{M}}^{-1} - \bar{\mathbf{M}}^{-1}\widetilde{\mathbf{M}}\left( \bar{\mathbf{M}}\right. \right. + \\
& \left. \left. +\widetilde{\mathbf{M}}\bar{\mathbf{M}}^{-1}\widetilde{\mathbf{M}}\right)^{-1}\widetilde{\mathbf{M}}\bar{\mathbf{M}}^{-1} \right\},
\end{split}
\label{CRB_IM_PART}
\end{equation}
where $\bar{\mathbf{M}} = Re\left\{ \mathbf{U}^{H}\mathbf{U} \right\}$ and $\widetilde{\mathbf{M}} = Im\{\mathbf{U^{H}}\mathbf{U}\}$.
}
\textcolor{black}{
Let us recall that $\mathbf{U} = \left( \mathbf{S} \otimes \mathbf{X} \otimes \mathbf{I}_L \right)$, $\mathbf{X}^{\textrm{H}}\mathbf{X} = T\mathbf{I}_M$, and $\mathbf{S}^{\textrm{H}}\mathbf{S} = K\mathbf{I}_N$. Hence, $\mathbf{U^{H}}\mathbf{U} = KT\mathbf{I}_{MNL}$. This implies that $\widetilde{\mathbf{M}} = \mathbf{0}$. The two above expressions can be simplified as }
\begin{eqnarray}
\textrm{CRB}(\bar{\boldsymbol{\theta}}) &=& \dfrac{\sigma^2}{2KT}\mathbf{I}_{MNL},\label{eq:crb1}\\
\textrm{CRB}(\widetilde{\boldsymbol{\theta}}) &=& \dfrac{\sigma^2}{2KT}\mathbf{I}_{MNL}.\label{eq:crb2}
\end{eqnarray}
Therefore, using definition (\ref{CRB_tot}):
\begin{eqnarray}
    \mathbb{E}\|\boldsymbol{\theta}_c - \boldsymbol{\hat{\theta}}_c\|^2 \geq \dfrac{\sigma^2}{KT}MNL.
\end{eqnarray}

It is important to note that there is no need to derive the mathematical expectation in the right-hand side of (\ref{CRB_tot}) over the parameters of interest and of nuisance due to the simple expression of the CRB.

\section{Simplified version of BALS}
\label{Appendix B}

Under the column-orthogonality assumption for $\mathbf{X}$ and $\mathbf{S}$, the right pseudo-inverses in (\ref{EstimaG}) and (\ref{EstimaH}) can be replaced by lower complexity matrix products, leading to a faster implementation of the BALS algorithm. Defining
$\mathbf{M}_1\doteq \mathbf{S} \diamond \mathbf{X}\mathbf{H}^{\textrm{T}}$ and $\mathbf{M}_2\doteq \mathbf{S} \diamond \mathbf{G}$, and
using property (\ref{Propertie Hadmard x Khatri}), we have
\begin{eqnarray}
\mathbf{M}_1^{\textrm{H}}\mathbf{M}_1&\hspace{-3ex}=& \hspace{-3ex} (\mathbf{S}^{\textrm{H}}\mathbf{S})  \odot (\mathbf{H}\mathbf{X}^{\textrm{H}}\mathbf{X}\mathbf{H}^{\textrm{H}})\nonumber\\
\hspace{-5ex}&=&\hspace{-1ex}
KT\left[\begin{array}{ccc} \hspace{-1ex}\|\mathbf{h}_1\|^2  & \hspace{-1ex}& \hspace{-1ex}\\\hspace{-1ex} & \hspace{-1ex}\ddots & \hspace{-1ex}\\ \hspace{-1ex}& \hspace{-1ex}& \hspace{-1ex}\|\mathbf{h}_N\|^2  \end{array}\right] \doteq KT\boldsymbol{\Sigma}_{\mathbf{H}}
\end{eqnarray}
and
\begin{eqnarray}
\mathbf{M}^{\textrm{H}}_2\mathbf{M}_2 \hspace{-1ex}&=& \hspace{-1ex} (\mathbf{S}^{\textrm{H}}\mathbf{S})  \odot (\mathbf{G}^{\textrm{H}}\mathbf{G})\nonumber\\
&=& \hspace{-1ex}
K\left[\begin{array}{ccc} \hspace{-1ex}\|\mathbf{g}_1\|^2  & \hspace{-1ex}&\hspace{-1ex} \\ \hspace{-1ex}& \hspace{-1ex}\ddots & \hspace{-1ex}\\ \hspace{-1ex}& \hspace{-1ex}& \hspace{-1ex}\|\mathbf{g}_N\|^2  \end{array}\right]\doteq K\boldsymbol{\Sigma}_{\mathbf{G}},
\end{eqnarray}
which implies that
\begin{eqnarray}
\hat{\mathbf{G}} &=& (1/KT)\cdot\mathbf{Y}_1\mathbf{M}_1^\ast\boldsymbol{\Sigma}^{-1}_{\mathbf{H}}\\
\vspace{2ex}
\hat{\mathbf{H}}^\textrm{T} &=& (1/KT)\cdot\mathbf{X}^{\textrm{H}}\mathbf{Y}_2\mathbf{M}_2^\ast\boldsymbol{\Sigma}^{-1}_{\mathbf{G}}.
\end{eqnarray}
Due to the diagonal structure of $\boldsymbol{\Sigma}_{\mathbf{H}}$  and $\boldsymbol{\Sigma}_{\mathbf{G}}$, these expressions provide lower complexity implementations of (\ref{EstimaG}) and (\ref{EstimaH}), respectively, by replacing matrix inversions by simpler matrix products. In particular, each update of $\hat{\mathbf{G}}$ and $\hat{\mathbf{H}}$ can be viewed as a set of $N$ independent processes (one for each IRS element) that can be carried out in parallel. The BALS is summarized in Algorithm \ref{PseudocodeBALS_With K>=N restriction}.
\begin{algorithm}[!t]
\IncMargin{1em}
	\DontPrintSemicolon
	\SetKwData{Left}{left}\SetKwData{This}{this}\SetKwData{Up}{up}
	\SetKwFunction{Union}{Union}\SetKwFunction{FindCompress}{FindCompress}
	\SetKwInOut{Input}{input}\SetKwInOut{Output}{output}
	\textbf{Procedure}\\
	\Input{$i = 0$; \textit{Initialize} $\hat{\mathbf{H}}_{(i=0)}$}
	\Output{$\hat{\mathbf{H}}$, $\hat{\mathbf{G}}$}
	\BlankLine
    \Begin{
	$i = i + 1 ;$\;
     \While{$\|e(i) - e(i-1)\| \geq \delta$}{
	    \begin{enumerate}
		\item [1:] \textit{Compute} $\hat{\mathbf{M}}_{1_{(i)}}=\mathbf{S} \diamond \mathbf{X}\hat{\mathbf{H}}_{(i-1)}$ \\ \textit{and find a least squares estimate of $\mathbf{G}$:}
		\vspace{0.1in}
		\begin{itemize}
		    \item[] $\hat{\mathbf{G}}_{(i)} = \dfrac{1}{KT}\mathbf{Y}_1\hat{\mathbf{M}}^{\ast}_{1_{(i)}}\boldsymbol{\hat{\Sigma}}^{-1}_{\mathbf{H}_{(i-1)}}$
		\end{itemize}
		 \vspace{0.1in}
		\item[2:] \textit{Compute} $\hat{\mathbf{M}}_{2_{(i)}}=\mathbf{S} \diamond \mathbf{X}\hat{\mathbf{G}}_{(i)}$ \\ \textit{and find a least squares estimate of $\mathbf{H}$:}
		\vspace{0.1in}
		\begin{itemize}
		    \item[] $\hat{\mathbf{H}}^{\textrm{T}}_{(i)} =\dfrac{1}{KT}\mathbf{X}^{\textrm{H}}\mathbf{Y}_2\hat{\mathbf{M}}^{\ast}_{2_{(i)}}\boldsymbol{\hat{\Sigma}}^{-1}_{\mathbf{G}_{(i)}}$
		\end{itemize}
		\vspace{0.1in}
		\item[3:]\textit{Repeat steps} $1$ to $2$ \textit{until convergence.}
	  \end{enumerate}
	  \textbf{end}
	  }
	  \textbf{end}
	  }
	\caption{Simplified BALS}
	\label{PseudocodeBALS_With K>=N restriction}
\end{algorithm}
\end{appendices}


\renewcommand{\baselinestretch}{.97}


\end{document}